\documentclass[fleqn,usenatbib]{mnras}

\usepackage{newtxtext,newtxmath}

\usepackage[T1]{fontenc}
\usepackage{ae,aecompl}
\usepackage{graphicx}
\usepackage{subfigure}
\usepackage{soul}

\newcommand{\sag}{\textsc{sag}}
\newcommand{\smdpl}{\textsc{smdpl}}

\usepackage{graphicx}	
\usepackage{amsmath}	
\usepackage{epsfig}

\usepackage[dvipsnames]{xcolor}

\usepackage[switch]{lineno}

\title[Environmental effects]%
{Environmental effects on associations of dwarf galaxies}
\author[Yaryura et al.]
{
C.Yamila~Yaryura$^{1,2}$\thanks{E-mail: yamila.yaryura@unc.edu.ar}, Mario G.~Abadi$^{1,2}$, Stefan~Gottl\"ober$^{3}$,
Noam I.~Libeskind$^{3,4}$, 
\newauthor  Sof\'ia A.~Cora$^{5,6}$, Andr\'es N.~Ruiz$^{1,2}$, Cristian A.~Vega-Mart\'{i}nez$^{7,8}$, 
Gustavo~Yepes $^{9,10}$
\\
\\
$^{1}$CONICET-Universidad Nacional de C\'{o}rdoba, Instituto de Astronom\'{i}a Te\'{o}rica y Experimental (IATE),
Laprida 854, X5000BGR, C\'{o}rdoba, Argentina\\
$^{2}$Observatorio Astron\'{o}mico, Universidad Nacional de C\'{o}rdoba, Laprida 854, X5000BGR, C\'{o}rdoba, Argentina\\
$^{3}$Leibniz-Institut f\"ur Astrophysik Potsdam (AIP), An der Sternwarte 16, D - 14482, Potsdam, Germany \\
$^{4}$University of Lyon; UCB Lyon 1/CNRS/IN2P3; IPN Lyon, France\\
$^{5}$Instituto de Astrof\'isica de La Plata (CCT La Plata, CONICET,UNLP), Observatorio Astron\'omico, Paseo del Bosque,\\ 1900FWA, La Plata, Argentina\\
$^{6}$Facultad de Ciencias Astron\'omicas y Geof\'{\i}sicas, Universidad Nacional de La Plata (UNLP), Observatorio Astron\'omico,\\ Paseo del Bosque, B1900FWA La Plata, Argentina\\
$^{7}$Instituto de Investigaci\'on Multidisciplinario en Ciencia y Tecnolog\'{\i}a, Universidad de La Serena, Ra\'ul Bitr\'an 1305, La Serena, Chile\\
$^{8}$Departamento de Astronom\'{\i}a, Universidad de La Serena, Av. Juan Cisternas 1200 Norte, La Serena, Chile.\\
$^{9}$Departamento de F\'{\i}sica Te\'orica M-8, Universidad Aut\'onoma de Madrid, Cantoblanco, E-28049 Madrid, Spain\\
$^{10}$Centro de Investigaci\'on Avanzada en F\'{\i}sica  Fundamental (CIAFF), Universidad Aut\'onoma de Madrid, E-28049 Madrid, Spain\\
}

\date{Accepted XXX. Received YYY; in original form ZZZ}

\pubyear{2022}

\begin{document}
\label{firstpage}
\pagerange{\pageref{firstpage}--\pageref{lastpage}}
\maketitle

\begin{abstract}
We study the properties of associations of dwarf galaxies and their dependence
on the environment. 
Associations of dwarf galaxies are extended systems
composed exclusively of dwarf galaxies, considering as dwarf galaxies those
galaxies less massive than \mbox{$M_{\star, \rm max} = 10^{9.0}$ ${\rm
M}_{\odot}\,h^{-1}$}.  
We identify these particular systems using a semi-analytical model of 
galaxy formation coupled to a dark matter only simulation in the 
$\Lambda$ Cold Dark Matter cosmological model. 
To classify the environment, we estimate eigenvalues from the tidal 
field of the dark matter particle distribution of the simulation.  
We find that the majority, two thirds, of associations are located in 
filaments ($ \sim 67$ per cent), followed by walls ($ \sim 26 $ per cent), 
while only a small fraction of them are in knots ($ \sim 6 $ per cent) 
and voids ($ \sim 1 $ per cent).
Associations located in more dense environments present significantly higher
velocity dispersion than those located in less dense environments, evidencing
that the environment plays a fundamental role in their dynamical properties.
However, this connection between velocity dispersion and the environment
depends exclusively on whether the systems are gravitational bound or unbound,
given that it disappears when we consider associations of dwarf galaxies that
are gravitationally bound.  
Although less than a dozen observationally detected associations of dwarf 
galaxies are currently known, our results are predictions on the eve of forthcoming 
large surveys of galaxies, which will enable these very particular systems 
to be identified and studied.
\end{abstract}




\section{Introduction} 

The large-scale structure map of the Universe reveals that galaxy 
and dark matter distributions are not uniform, describing an intricate 
interconnected network known as the cosmic web.
Within this network, galaxies, intergalactic gas, and dark matter 
are distributed within high density regions
as groups and clusters of galaxies, or along 
filaments and sheet-like walls, which surround very low density 
regions known as voids.
Most of the galaxies embedded in this cosmic web belong to systems 
which can contain from a few to hundreds even thousands 
of members \citep{H&G:1982,Yang:2007,Tempel:2012}.
These systems have been extensively studied
and there is ample evidence that many of their observed 
properties are influenced by the web-like environment 
\citep[][among others]{Dressler:1980,Kauffmann:2004,OMill:2008,Peng:2010,Zheng:2017,Duplancic:2020}.
For example, it is known that elliptical galaxies are located 
more frequently in denser regions, while spiral galaxies are more common 
in the field \citep{Dressler:1980}.
Similar trends are also detected for colours, star formation history 
and the ages of galaxies \citep{Blanton:2005}; in denser environments, 
there is a higher proportion of red and passive galaxies for a 
given stellar mass \citep{Wetzel2012, Wang2018}.
On the other hand, from a theoretical point of view, many studies 
show that the orientation of the haloes' minor axes shows a tendency to 
be perpendicular to the wall or filament where they reside. 
The spin orientation also correlates with the halo mass, being 
parallel to the filaments or walls for low-mass haloes and 
perpendicular for higher mass haloes 
\citep[][among others]{Aragon:2007,Hahn:2007a,Hahn:2007b,Zhang:2009,Libeskind2013}.
\\

Groups of galaxies are a particular type of system, 
being the most common structures of galaxies in the Universe.
Even though there is no clear demarcation between 
groups and clusters, the latest one are generally considered to contain 
hundreds or thousands of galaxies while groups contain only a few, 
with $\sim 50$ being the most commonly used cut-off value when defining them.
Their typical sizes are on average compared to 
a spherical volume of $\sim 1 \,{\rm Mpc}$ of diameter. 
Their virial masses are, on average, 
approximately $\sim 10^{13} \,{\rm  M}_{\odot}\,h^{-1}$ 
and the velocity dispersion of their galaxy members
are about $\sim 150 \,{\rm km\,s^{-1}}$.
These groups can host very bright galaxies as well as fainter galaxies. 
However, the pioneering work of \cite{Tully:1987} revealed 
the existence of a very striking type of groups called 
\textit{`associations of dwarf galaxies'}. 
These systems have the particularity of being extended systems
, with typical sizes of $\sim 0.2\,{\rm Mpc}\,h^{-1}$,
composed only of dwarf galaxies, extracted from the 
Nearby Galaxies Catalog \citep{Tully:1988}.
They use a merging-tree algorithm to define these systems, 
where the luminosity density, determined by the combined luminosities and 
separations of contributing systems, was used to characterize the linkages 
between galaxies. 
Two levels of structure, namely `groups' and `associations' were defined 
based on luminosity density thresholds. 
We focus on \textit{`associations of dwarf galaxies'} derived from linkages 
between galaxies that had insignificant luminosities, making the luminosity 
density fail to reach the threshold required to be classified as a group 
(see \citealt{Tully:1987} for a exhaustive depiction of the method).
\\

Among the very few works that study these 
particular systems, \cite{Tully:2006} 
present a detailed description of 
the only seven associations of dwarf galaxies observed up to now and 
their main dynamical properties.
Among these properties we can highlight their velocity dispersions covering  
a range between $\sim 20$ and $\sim 75 \,{\rm km\,s^{-1}}$, their 
sizes around $0.2 \,{\rm Mpc}\,h^{-1}$ and their virial masses 
ranging between $\sim 10^{10.5}$ and 
$\sim 10^{11.8} \,{\rm  M}_{\odot}\,h^{-1}$.
From the theoretical point of view, \cite{Yaryura:2020} present a study 
of associations of dwarf galaxies in the cosmological framework of 
the $\Lambda$ Cold Dark Matter ($\Lambda$CDM) model, applying a semi-analytic 
model of galaxy formation to a dark matter-only {\em N}-body numerical simulation. 
They conclude that the $\Lambda$CDM model is able to reproduce these particular 
systems.
On average, these simulated systems have typical sizes of $\sim 0.2\,{\rm Mpc}\,h^{-1}$, 
velocity dispersions of $\sim 30 \,{\rm km\,s^{-1}} $ and estimated total 
masses of $\sim 10^{11} \,{\rm  M}_{\odot}\,h^{-1}$. 
These main dynamical properties mean values are 
comparable to the observational results presented by \cite{Tully:2006}.
In comparison with groups of galaxies, these associations present 
lower masses and velocity dispersions despite their large size.
Their low masses, in addition to their 
low luminosity, suggest that their 
mass–to–light ratio is relatively high if these systems are 
bound systems. 
Based on this assertion, \cite{Tully:2006} infer that these associations 
are bound but dynamically unevolved systems.
They also suggest that they presumably contain dark matter 
subhaloes ranging from $10^9$ to $10^{10}$ $M_{\odot}$ which 
contain insufficient amounts of gas and stars to be detected at present.
\\

So far only a few of these associations have been 
observed and very little is known about their properties. 
But these systems are of fundamental importance because they could 
be used as a  probe of the cosmological model, if a substantial number 
of them will be detected in future surveys. 
Currently, the standard paradigm, $\Lambda$CDM, is a theory predicting 
evolution of haloes due to mergers. 
In fact,  at virtually any given moment  in cosmic time, dark matter haloes 
undergo mergers. 
The relevant time scales here are well known. 
Therefore, it is intriguing to examine the nature of associations 
that withstand the cosmic forces of tides and gravity.
\\

From the observational point of view, although currently 
only less than a dozen of these systems of dwarf galaxies are known, 
their study holds significant importance in anticipation of upcoming 
galaxy surveys, such as 
the Dark Energy Spectroscopic Instrument (DESI\footnote{\url{https://www.desi.lbl.gov/}}), 
the Vera C. Rubin Observatory (\citealt{LSST:2019}), 
the 4-metre Multi-Object Spectroscopic Telescope 
(4MOST, \citealt{deJong:2019}), among many others.
These surveys hold the promise of providing a highly detailed map of the 
Local Universe, facilitating the detection of faint galaxies that have 
remained elusive until now. 
By detecting these faint galaxies, we anticipate the possibility of 
identifying new systems exclusively composed of dwarf galaxies. 
In this sense, our findings will consist of theoretical predictions 
that await observational confirmation when these future galaxy catalogs 
become available.
\\

The main goal of this paper is to deepen the theoretical understanding of 
these associations of dwarf galaxies by analysing the large-scale 
environment within which they form and evolve, and to study how their
main dynamical properties vary with the environment.
For this, we use the semi-analytic model of galaxy formation 
\sag~\citep[acronym for Semi-Analytic Galaxies,][]{Cora:2018} coupled to 
the $400\,h^{-1} \,{\rm Mpc}$ \textsc{Small MultiDark Planck} simulation 
(\smdpl) based on the Planck cosmology \citep{Klypin:2016}. 
\smdpl~simulation is publicly available in the \textsc{CosmoSim} database 
\footnote{\url{https://www.cosmosim.org}}.
\\

This paper is organized as follows. 
We describe the \smdpl~simulation and the \sag~model in Section \ref{S_methods}. 
In Section \ref{S_asso}, we define our sample of associations of dwarf galaxies 
and describe their main properties.
Section \ref{S_env} classifies the large-scale environment 
and analyses the dependence of the properties of the associations on the environment.
In Section \ref{S_conclusions}, we summarize our main 
results and present our conclusions.
\\

\begin{table}
  \centering
  \begin{tabular}{l r r}
    \hline
    Parameter &  Best-fitting value\\
    \hline
    $\alpha$           & 0.08                    \\
    $\epsilon$         & 0.53                    \\
    $\epsilon_{\rm ejec}$  & 0.01               \\
    $f_{\rm BH}$       & 0.07                    \\
    $\kappa_{\rm AGN}$ & 1.18 $\times$ 10$^{-5}$ \\
    $f_{\rm pert}$     & 31.22                    \\
    $\gamma$      & 0.005                    \\
    $f_{\rm hot,sat}$      & 0.26                    \\
    \hline                                                                
  \end{tabular}
  \caption{Best-fitting values of the free parameters of \sag~model obtained 
  with the~PSO technique. 
  This set of values is obtained from the application of \sag~to the merger 
  trees of the subbox selected from the \textsc{SMDPL} simulation.}
      \label{table:parameters}
\end{table}

\section{Hybrid model of galaxy formation} \label{S_methods}

The sample of associations of dwarf galaxies is extracted from a galaxy 
catalogue constructed by applying a hybrid model of galaxy formation that 
couples a semi-analytical model of galaxy formation and evolution 
with a dark matter-only cosmological simulation. 
Below, we briefly describe the main aspects of this model.
\\

\subsection{Dark matter cosmological simulation}
\label{sec:simSMDPL}

We use the \smdpl~dark matter-only cosmological
simulation\footnote{doi:10.17876/cosmosim/smdpl/.}, which follows the 
evolution of $3840^3$ particles from redshift $z = 120$ to $z = 0$, 
within a (relatively) small volume (a periodic box 
of side-length of $400\,{\rm Mpc}\,h^{-1}$).
This large number of particles within such a volume 
reaches a mass resolution 
of $9.63\times10^7 \,{\rm M}_{\odot}\,h^{-1}$ per dark matter (DM) 
particle (see \citealt{Klypin:2016} for more details).
\smdpl~cosmological parameters are given by a flat $\Lambda$CDM model 
consistent with Planck measurements: 
\mbox{$\Omega_{\rm m}$ = 0.307}, 
\mbox{$\Omega_{\rm B}$ = 0.048}, \mbox{$\Omega_{\Lambda}$ = 0.693}, 
\mbox{$\sigma_{8}$ = 0.829},
\mbox{$n_{\rm s}$ = 0.96} and \mbox{$h$ = 0.678}, \citep{Planck:2014}.
\\

The \textsc{Rockstar} halo finder \citep{Behroozi_rockstar} is used 
to identify DM haloes keeping just overdensities with at 
least $N_\text{min}$~=~20~DM particles. 
There are two classifications of DM haloes: \textsl{main host} haloes
(detected over the background density) and \textsl{subhaloes} 
(that lie inside other DM haloes). 
From these haloes, \textsc{ConsistentTrees} \citep{Behroozi_ctrees} 
constructs merger trees, by linking haloes and subhaloes forwards 
and backwards in time to progenitors and descendants, respectively. 
\\

\subsection{Semi-analytic model of galaxy formation SAG}

In this project, we follow \cite{Yaryura:2020} and use the latest 
version of the semi-analytic model \sag~presented in \cite{Cora:2018}, 
based on the model previously presented by \cite{Springel:2001}. 
This is an updated and improved version, including the main physical 
processes required for galaxy formation: gas cooling, star formation 
in quiescent and bursty modes (being the latter triggered by 
disc instabilities and mergers), black hole growth, feedback from 
supernovae and active galactic nuclei (AGN), environmental effects 
(ram pressure stripping and tidal stripping), chemical enrichment. 
The circulation of mass and metals among the different baryonic components 
(hot gas halo, cold gas disc, stellar disc and bulge) are regulated by 
ejection and reincorporation mechanisms associated to feedback 
processes and recycling of stellar mass. We refer the reader to 
\cite{Cora:2018} for a detailed and exhaustive description of the model.
\\

Each \sag~galaxy populates a DM halo of the simulation in 
such a way that central galaxies correspond to
main host haloes while satellite galaxies are hosted by subhaloes, 
according to the information provided by the merger trees.
A satellite galaxy is defined as an orphan galaxy when its 
DM substructure is no longer detected by the halo finder.
The position and velocity of orphan galaxies are obtained following 
the model introduced by \citet{Delfino2022} to calculate the orbital 
evolution of unresolved subhaloes\footnote{In this work, 
we use a previous version of the orbital evolution code, where an isothermal 
sphere models the mass profile of both the host halo and unresolved subhaloes.}.
\\

The main galaxy properties provided by \sag~are 
listed in \citet[][see their Table A2]{Knebe:2018}, although the information 
of many other properties can be obtained as requested by a given project. 
The properties used in the current work are: 
a pointer to the DM halo in which a galaxy orbits; galaxy type 
(central galaxy, satellite with DM substructure, orphan satellite);
positions and velocities of a galaxy; stellar mass of a galaxy, $M_{\star}$; 
mass of the main host DM halo in which a galaxy resides, $M_{200}$. 
The halo mass is defined as the mass enclosed by a sphere of radius $r_{200}$, 
within which the mean density is a factor $\Delta=200$ times the critical 
density of the Universe $\rho_\textrm{c}$, i.e.,

\begin{equation}
   M_\textrm{200}(<r_\textrm{200}) = \Delta \rho_\textrm{c}
            \frac{4 \pi}{3} r_\textrm{200}^3.
   \label{eq:Mvir}
\end{equation}

To regulate the physical processes involved in the \sag~model, a set of 
free parameters are employed: the star formation efficiency ($\alpha$); the
efficiency of SN feedback from stars formed in both the disc and the bulge
($\epsilon$); the efficiency of ejection of gas from the
hot phase ($\epsilon_\text{ejec}$) and of its
reincorporation ($\gamma$); the growth
of super massive black holes and efficiency of AGN feedback ($f_\text{BH}$ and
$\kappa_\text{AGN}$, respectively); the factor involved in the distance scale
of perturbation to trigger disc instability events ($f_{\rm pert}$); and the
fraction that determines the destination of the reheated cold gas of a satellite galaxy
($f_\text{hot,sat}$; when the hot gas mass of a satellite
drops below a fraction $f_\text{hot,sat}$ of its baryonic mass, 
the reheated mass and associated metals are transferred to the corresponding 
central galaxy instead of being transported to the satellite's hot gas halo).
The parameter that regulates the redshift dependence of the SNe feedback 
was not allowed to vary during the calibration process but fixed in $1.3$ 
according to the fit found by \citet{Muratov:2015} from the analysis of 
their cosmological hydrodynamical simulations; this value allows \sag~to 
provide stellar mass and halo mass dependencies of the fractions of local 
quenched galaxies in better agreement with observational 
data \citep[][see their fig. 11]{Cora:2018}.
\\

These parameters are calibrated using a set of observed galaxy
properties:
i) the stellar mass functions at $z=0$ and $z=2$, for which we 
adopt the compilation data used by \citet{henriques_mcmc_2015}; 
ii) the star formation rate distribution function, which is the 
number density of galaxies in a certain interval of star formation rate; 
in this case, we use data from a flux-limited sample of galaxies 
observed with the \textsl{Herschel} satellite in the redshift 
range $z \in [0.0,0.3]$ \citep{gruppioni15}; 
iii) the fraction of mass in cold gas as a function of stellar mass; 
iv) the relation between bulge mass and the mass of the central
supermassive black hole (BH).
For these two latter relationships, we adopt observational 
data from \citet{boselli14}, which is based on a volume 
limited sample, within the range 
$\rm{log}(M_{\star}[{\rm M}_{\odot}]) \in [9.15,10.52]$ 
in stellar mass, and a combination of the datasets from
\cite{mcconnell_bhb_2013} and \cite{kormendy_bhb_2013}, respectively.
The best-fitting values of the free parameters of \sag~for 
the \smdpl~simulation were selected using the Particle 
Swarm Optimization technique \citep[PSO,][]{Ruiz:2015}, which are 
presented in Table~\ref{table:parameters}.

\begin{figure}
    \begin{subfigure}{
         \includegraphics[width=\columnwidth]
         {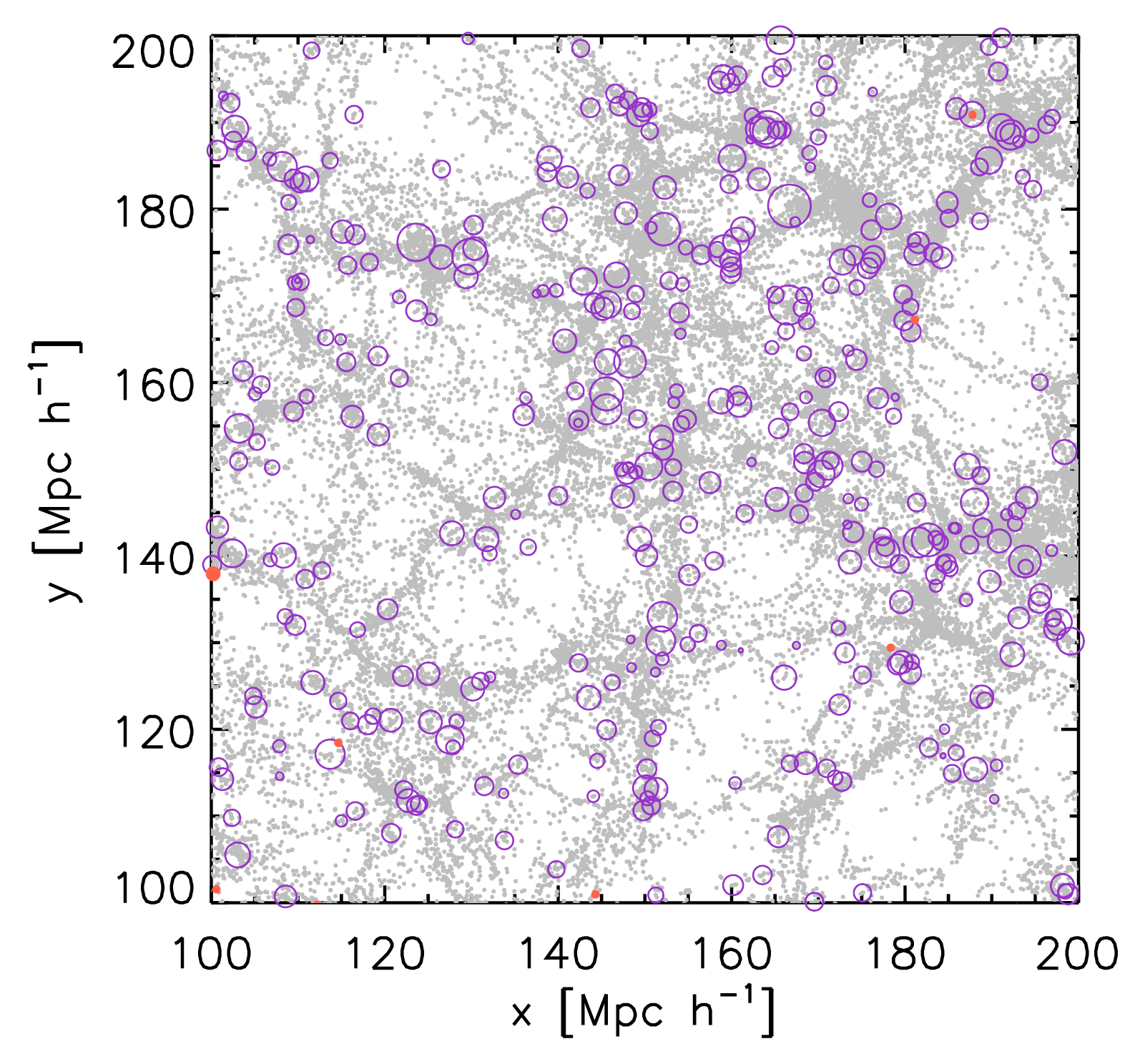}}
     \end{subfigure}
    \caption{Spatial projected distribution of our semi-analytical galaxies 
   (grey small dots) in a slice of $10\,{\rm Mpc}\,h^{-1}$ thickness. 
   Purple empty circles show the position 
   of the centre of mass of systems identified for the sample 
   {\em Dwarf\_associations}, while orange filled circles correspond to 
   systems identified for the sample {\em Dwarf\_groups}. 
   Systems in {\em Dwarf\_associations} dominate in number and 
   there are very few systems in 
   {\em Dwarf\_groups}, so it is difficult to see them. 
   For both samples, circle radius indicates $5 R_{I}$, where $R_{I}$ is the 
   inertial radius defined in equation~(\ref{eq:radius}). 
   We do not show systems of {\em Dwarf\_mix} sample to make the plot clearer.}
    \label{fig:pos}
\end{figure}

\begin{table*}
\centering
   \begin{tabular}{  c  c  c  c  c  c  c  c  c }
    \hline
    \hline
    \\
       &\multicolumn{1}{c}{$M_{\star} \text{max} [M_{\odot}/h]$} & \multicolumn{1}{c}{Main dark matter halo} & \multicolumn{1}{c}{$N_\text{systems}$} & \multicolumn{1}{c}{N=3} & \multicolumn{1}{c}{N=4} & \multicolumn{1}{c}{$N\ge5$} &
    \\
    \hline
    \textbf{\em Dwarf\_associations} & $10^{9.0}$ & different &  308250 & 211243 & 61849 & 35157 &
    \\
    \textbf{\em Dwarf\_mix} & $10^{9.0}$ & mix & 257277 & 125006 & 66487 & 65783 &
    \\
    \textbf{\em Dwarf\_groups} & $10^{9.0}$ & same &  40789 & 35243 & 4622 & 923 &
    \\
    \textbf{\em All\_associations} & $\infty$ & different & 429444 & 275708 & 90465 & 63270 &
     \\
     \textbf{\em All\_mix} & $\infty$ & mix & 890712 & 220785 & 163809 & 506117 &
     \\
    \textbf{\em All\_groups} & $\infty$ & same & 153368 & 99957 & 30247 & 23163 &
     \\
    \hline
    \hline
  \end{tabular}
  \caption{Different samples analysed throughout this paper, described in detail in Section \ref{S_asso_1}.
  First column shows the name of the sample.
  Second column indicates the maximum stellar mass threshold. 
  Third column indicates if galaxy members belong to the same main host halo, 
  to different ones or to a mix of both previous situations.
  Fourth column shows the number of systems in each sample. 
  Fifth, sixth and seventh columns indicate the number of systems with 3,  4 and 5 or more members, respectively.}
 \label{table:samples}
\end{table*}

\section{Associations of Dwarf Galaxies} \label{S_asso}

\subsection{Samples} \label{S_asso_1}

\begin{figure*}
    \begin{subfigure}{
       \includegraphics[width=0.5\textwidth]
       {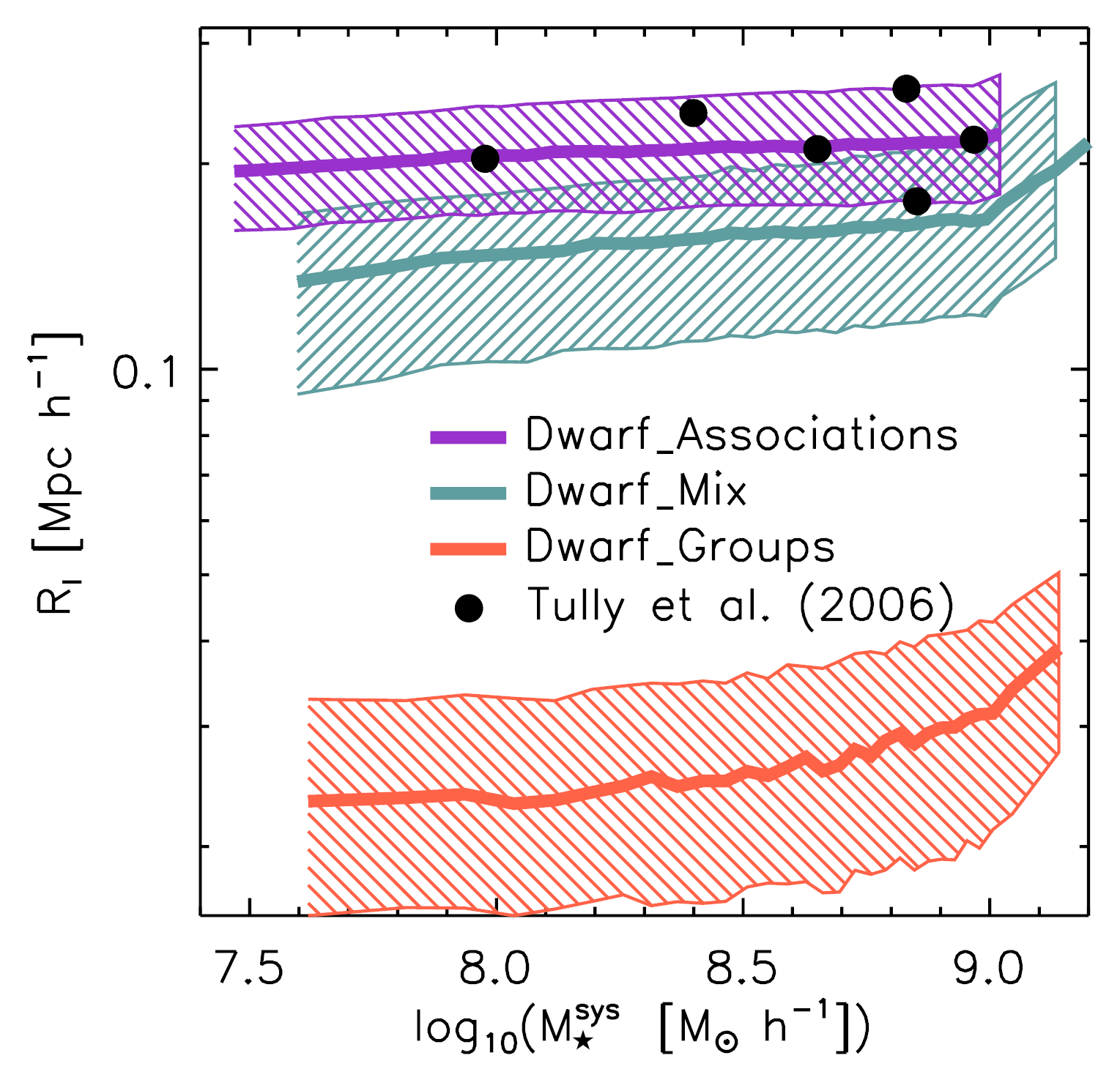}
       \includegraphics[width=0.5\textwidth] 
       {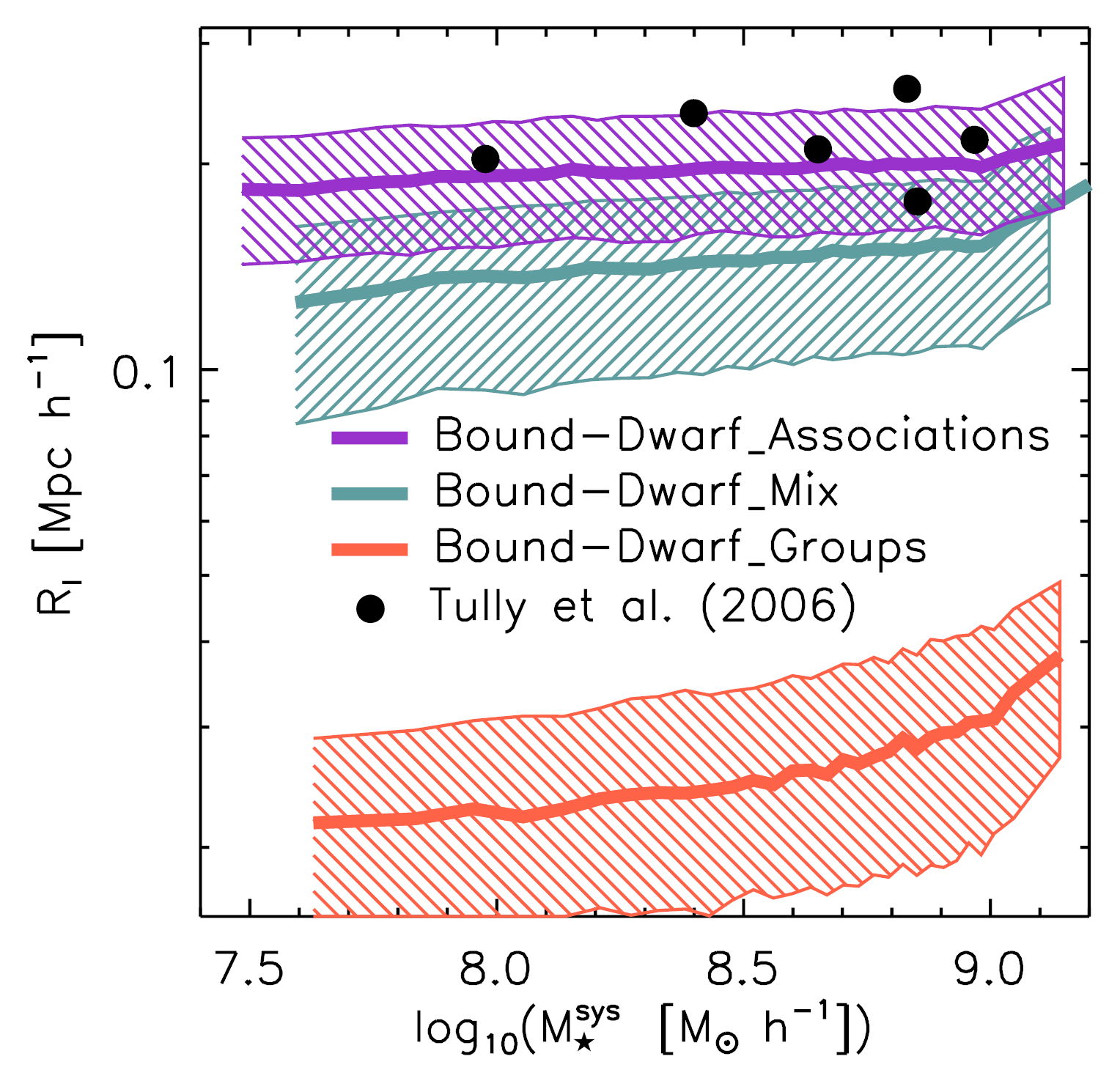}}
    \end{subfigure}
    \caption{Left panel: Size of dwarf galaxy systems ($R_{I}$) as a function of the 
    stellar mass of the system ($M_{\star}^{\rm sys}$). 
    Different coloured lines correspond to different samples: {\em Dwarf\_associations} 
    (purple line), {\em Dwarf\_mix} (green line) and {\em Dwarf\_groups} (orange line). 
    Solid lines show median values, taking equal number of bins in $M^{\rm sys}_{\star}$. 
    Shaded regions cover from $25$ per cent to $75$ per cent for each sample. 
    They are compared with observational results for dwarf galaxy associations taken from 
    \citet[][black filled circles]{Tully:2006}, assuming for them a factor equal 
    to one in the mass-to-light ratio. 
    Rigth panel: Idem as left panel but considering sub--samples of only 
    gravitationally bound systems: {\em Bound--Dwarf\_associations} (purple line), 
    {\em Bound--Dwarf\_mix} (green line) and {\em Bound--Dwarf\_groups} (orange line).}
    \label{fig:size}
\end{figure*}

The galaxy samples analyzed in this work are obtained from a 
set of semi-analytical galaxies built by
enforcing a minimum stellar and halo mass of  
$M_{\star} = 10^{6.8}\,{\rm M}_{\odot}\,h^{-1}$, and 
$M_{200} = 10^{9.28} \, {\rm M}_{\odot}\,h^{-1}$ 
(equivalent to $20$ DM particles), respectively.
The final sample has a total of $26\,506\,948$ well resolved, 
semi-analytical galaxies, with stellar masses ranging between 
$6.8 < {\rm log_{10}}(M_{\star}[{\rm M}_{\odot}\,h^{-1}]) < 12.9$
and halo masses ranging between 
$9.28 < {\rm log_{10}}(M_{200}[{\rm M}_{\odot}\,h^{-1}]) < 15.17$.
A well-established percolation algorithm, known as {\em friends-of-friends} 
\citep[FOF,][]{H&G:1982}, is used to identify galaxy systems with 
sizes similar to the observed associations presented in \citet{Tully:2006}. 
To define our samples, we follow the procedure described in \cite{Yaryura:2020}.
We use a linking length of $0.4 \,{\rm Mpc}\,h^{-1}$ and select 
systems of galaxies with at least 3 members. 
We select this linking length value following \cite{Yaryura:2020} 
who analyze characteristic sizes of systems identified varying the linking 
length parameter between $0.3 \,{\rm Mpc}\,h^{-1}$ and $0.5 \,{\rm Mpc}\,h^{-1}$. 
Characteristic sizes of the systems are sensitive to the chosen linking length, 
being more extended when they use a greater linking length value.
Comparing with the observational results presented by Tully et al (2006), 
they conclude that $0.4 \,{\rm Mpc}\,h^{-1}$ is the best choice for the analysis 
of these associations.
In a first instance, we do not include the velocity of 
the systems as a selection criteria since our main objective is to 
mimic observation. 
We want to identify systems with properties comparable to the observations, 
but considering the fewest possible restrictions in their selection.
\\

As we are interested in systems made up only of dwarf galaxies, 
we remove those systems that have one (or more) 
galaxy member more massive than a stellar mass threshold $M_{\star, \rm max}$. 
In this paper, we consider dwarf galaxies as those galaxies less 
massive than \mbox{$M_{\star, \rm max} = 10^{9.0}$ ${\rm  M}_{\odot}\,h^{-1}$}
(we check that our results do not vary significantly if we consider a more restrictive value 
as \mbox{$M_{\star, \rm max} = 10^{8.5}$ ${\rm  M}_{\odot}\,h^{-1}$}, or a less restrictive value 
as \mbox{$M_{\star, \rm max} = 10^{9.5}$ ${\rm  M}_{\odot}\,h^{-1}$}).
Once we selected systems made up only of dwarf galaxies, 
and following the classification made by \cite{Yaryura:2020},
we split the sample into three different sub-samples according to the following conditions: 
(i) systems with all their galaxy members belonging to the same main DM halo;
(ii)  systems with all their galaxy members belonging to different main host DM haloes;
(iii) systems for which some of the galaxies belong to the same main DM halo, but others 
belong to different main host haloes. 
We refer to these respective sub-samples as: 
(i) {\em Dwarf\_groups}, (ii) {\em Dwarf\_associations}, and (iii) {\em Dwarf\_mix}. 
It is important to note that each system belongs to only 
one of these sub-samples. 
For comparison, we also consider samples without restriction in the 
maximum value of the stellar mass of member galaxies.
In those cases, the samples are called {\em All\_groups}, {\em All\_associations}, 
and {\em All\_mix}, according to the previously described criterion of belonging 
to a DM halo.
Table \ref{table:samples} summarises these samples. 
\\

Fig.~\ref{fig:pos} shows the spatial projected distribution of 
semi-analytical galaxies (small grey dots) in a slice of 
$10\,{\rm Mpc}\,h^{-1}$ thickness.
Open purple circles show the position 
of the systems in the sample {\em Dwarf\_associations}, while 
orange filled circles correspond to systems identified 
in the sample {\em Dwarf\_groups}.
To make the plot clearer, the circle radius indicates $5 R_{\rm I}$, 
where $R_{\rm I}$ is the inertial radius considered as an indicator 
of the size of the system, defined by

\begin{equation}
R_{\rm I} = \left( \sum_{i}^{N} r_{i}^{2}/N \right) ^{1/2},
\label{eq:radius}
\end{equation}

\noindent where $r_{i}$ is the three--dimensional distance of a galaxy from the 
system centroid and the sum for each system is performed over all members ($N$). 
From this plot, it is evident that associations of dwarf galaxies, with all 
their galaxy members belonging to different main host DM haloes are much 
more numerous and noticeably larger than systems with all their galaxy 
members belonging to the same main host DM halo. 
It is expected that {\em Dwarf\_associations} are more numerous 
than {\em Dwarf\_groups} due to the resolution limit. 
Requiring a minimum halo mass of 
$M_{200} = 10^{9.28} \, {\rm M}_{\odot}\,h^{-1}$, 
there are large number of dwarf galaxies that are not resolved and 
therefore are not taken into account. 
These ignored dwarf galaxies, which accompany the central galaxy of the 
main host halo, could form groups of dwarf galaxies that are dismissed 
in our analysis.
However, although resolution effects can affect differently groups and 
associations, their relative abundance is not the main subject of our paper 
which focalizes in their dynamical and environmental properties.
\\

\begin{figure}
    \begin{subfigure}{
       \includegraphics[width=\columnwidth]{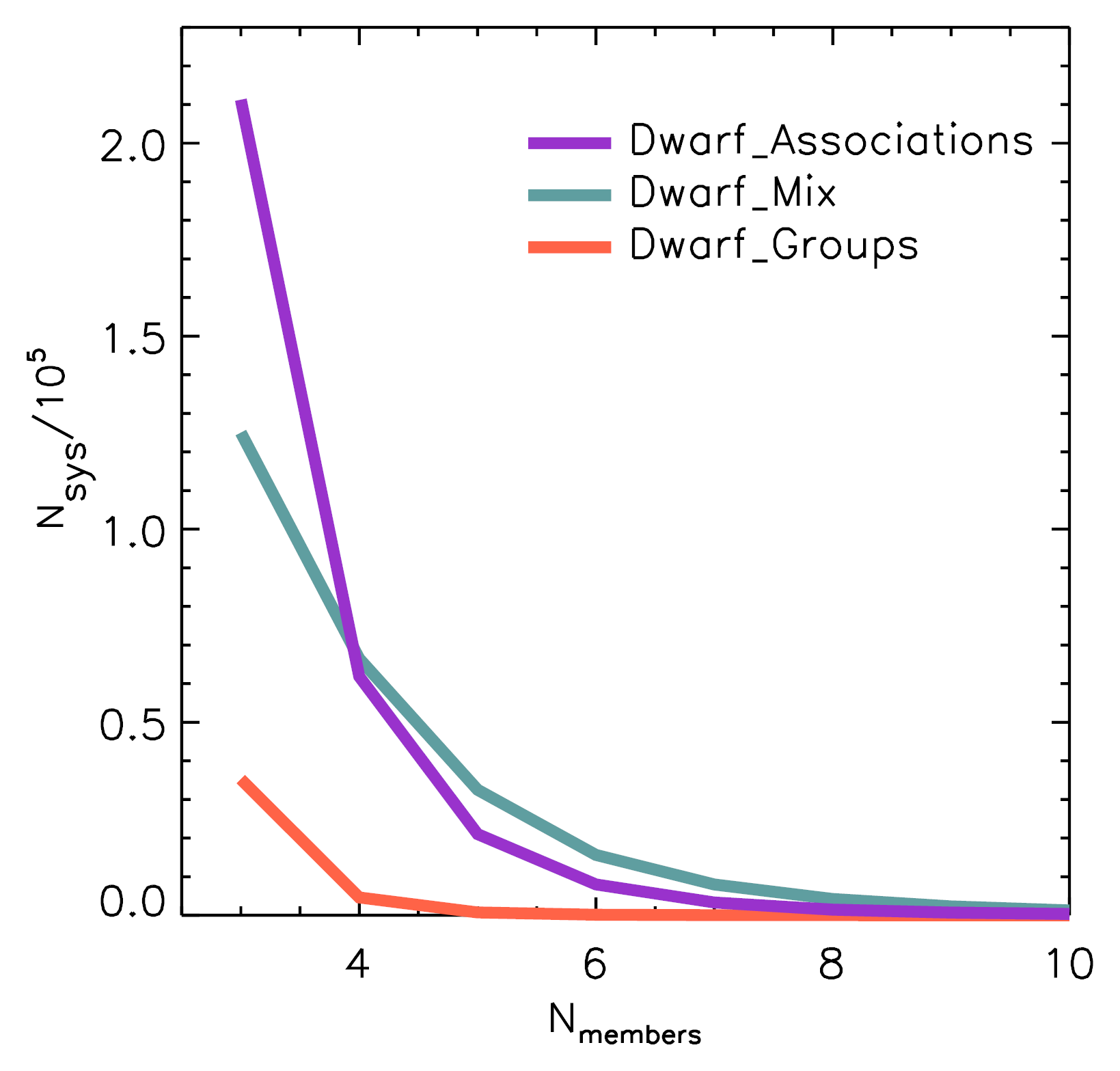}}
    \end{subfigure}
    \caption{Number of galaxy systems as a function of the 
    number of members per system. 
    Different coloured lines correspond to different samples: {\em Dwarf\_associations} 
    (purple line), {\em Dwarf\_mix} (green line) and {\em Dwarf\_groups} (orange line).}
    \label{fig:nm}
\end{figure}

To identify which of the previously defined samples most accurately reproduces 
the observed data, we analyse different dynamical properties and attempt to match 
the observed size-stellar mass relation from \citet{Tully:2006}.
Left panel of Fig.~\ref{fig:size} shows the size of the systems ($R_{\rm I}$, 
defined in equation~(\ref{eq:radius})) as a function of 
the stellar mass of the system ($M_{\star}^{\rm sys}$), defined as the 
sum of the galaxy stellar mass of all the galaxies defining each 
association or group.
For comparison, we also plot the size of the observed dwarf galaxy 
associations taken from \cite{Tully:2006} (black filled circles). 
This comparison is based on the {\em B}-band luminosity, $L_{B}$, of 
these observed systems, assuming a mass--luminosity ratio equal to 1. 
Notice that assuming a different value for the mass--luminosity ratio would only 
cause a horizontal shift in our results, which does not modify the conclusions. 
Systems in the sample {\em Dwarf\_associations} do much better 
in reproducing the empirical 
characteristic sizes of the observational sample of dwarf galaxies 
associations, in agreement with the results presented by \cite{Yaryura:2020}.
From this figure, it is evident that systems in the sample 
{\em Dwarf\_groups} are systematically smaller than systems in 
the sample {\em Dwarf\_associations} by a factor of $\sim 5$.
\\

Associations presented by \cite{Tully:2006} are 
systems identified from the spatial distribution of dwarf galaxies, so  
we do not have information regarding the full dynamics of these systems.
As a next step and taking advantage of the information provided by simulations, 
we estimate the binding energy of our theoretical systems compound only by 
dwarf galaxies to deduce whether they are gravitationally bound.
Then, we analyze subsamples of previously described samples 
({\em Dwarf\_associations}, {\em Dwarf\_mix} and {\em Dwarf\_groups})
considering only gravitationally bound systems. 
We estimate the binding energy of the systems by considering the contribution 
of each member galaxy, classifying a system as gravitationally bound if 
its binding energy is negative.
For consistency, we name these sub--samples as {\em Bound--Dwarf\_associations} 
(96902 systems, $\sim 31$ per cent of the original sample {\em Dwarf\_associations}), 
{\em Bound--Dwarf\_mix} (181046 systems, $\sim 70$ per cent of the original sample) 
and {\em Bound--Dwarf\_groups} (39048 systems, $\sim 96$ per cent of the original sample). 
Right panel of Fig.~\ref{fig:size} shows same as left panel but 
considering sub--samples of only gravitationally bound systems.
There are no significant differences between both panels, indicating 
that the sample of {\em Bound--Dwarf\_associations} 
show more similar results to the observational sample of dwarf galaxies 
associations, same as left panel.
\\
\\
It is evident that observed associations have characteristic sizes 
comparable with {\em Dwarf\_associations}, from where we could infer that 
member galaxies of the observed associations would be located in different 
main DM haloes. 
Furthermore, if the observed associations were in the same DM halo, 
it would likely be a large halo (due to the extended size of these systems), 
making it unlikely that the central galaxy is a dwarf galaxy.
Therefore, based on Fig.~\ref{fig:size}, henceforth we concentrate on 
the samples {\em Dwarf\_associations} and its sub--sample 
{\em Bound--Dwarf\_associations} to analyse how the 
environment affects their main dynamical properties. 
The rest of the samples are used for comparison.
\\ 
 
\begin{figure}
    \begin{subfigure}{
        \includegraphics[width=\columnwidth]{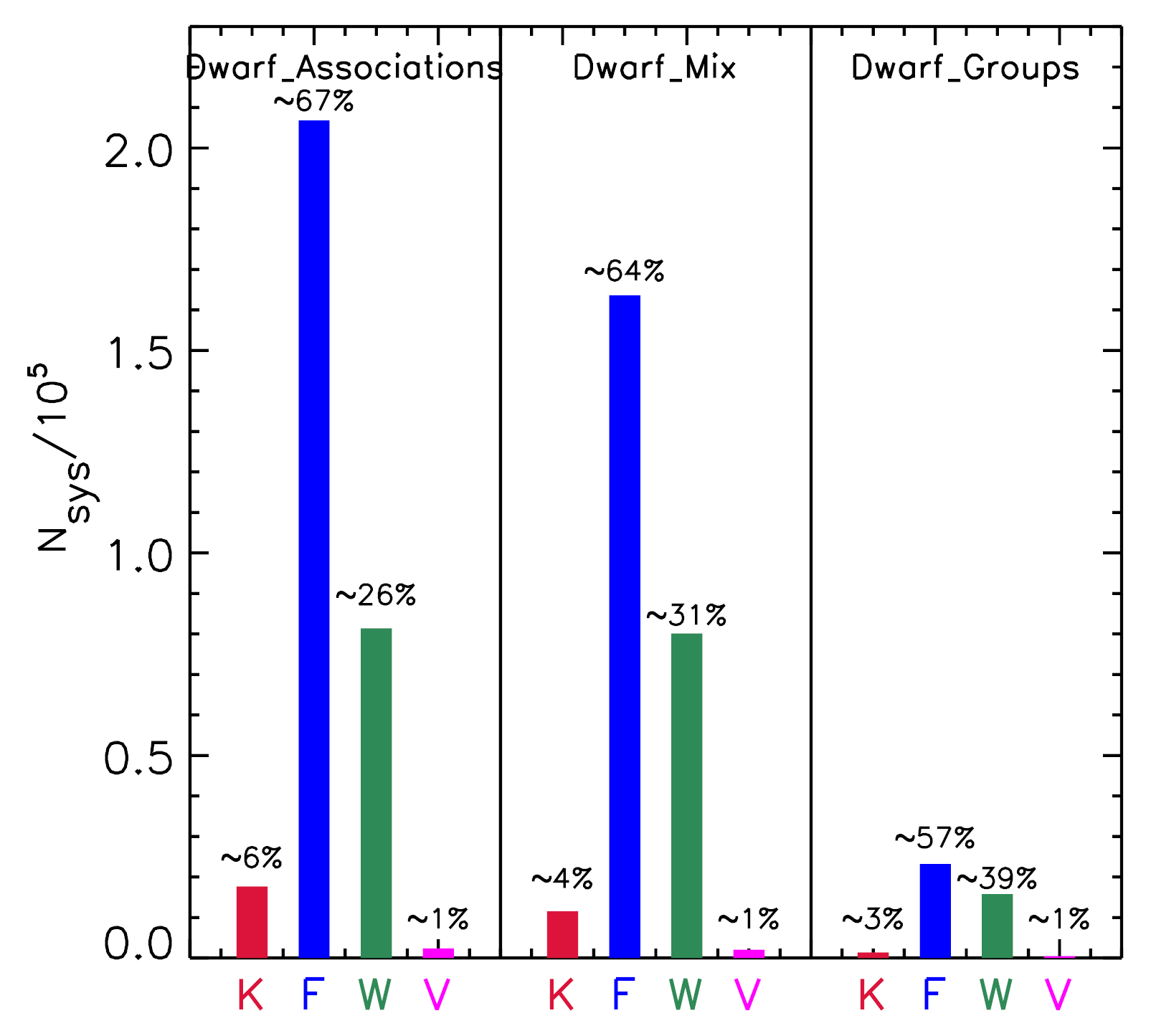}}
    \end{subfigure}
    \caption{Number of systems located in each 
    large-scale environment classification: knot (red bars), 
    filament (blue bars), wall (green bars) and void (magenta bars);
    for the samples {\em Dwarf\_associations} (left panel), 
    {\em Dwarf\_mix} (middle panel) and {\em Dwarf\_groups} (right panel). 
    Percentage values with respect to the total 
    of each sample are indicated above each bar.}
    \label{fig:histo_KFWV}
\end{figure}

To better understand whether galaxy members of {\em Dwarf\_associations} 
are preferably central galaxies, satellite galaxies with DM 
substructure or orphan galaxies, we study the internal 
structure of these associations. 
We find that $90$ per cent of these systems are composed 
of 3 or 4 members, most of which are central
galaxies ($\sim 97.5$ per cent) while the rest are orphan 
galaxies ($\sim 2.5$ per cent).
Fig.~\ref{fig:nm} shows the number of galaxy systems 
as a function of the number of members per system. 
Different coloured lines correspond to different samples: 
{\em Dwarf\_associations} (purple line), {\em Dwarf\_mix} (green line) 
and {\em Dwarf\_groups} (orange line). 
From this figure it is evident that most of the systems are composed 
just of 3 or 4 members: $90$ per cent for {\em Dwarf\_associations}, 
$75$ per cent for {\em Dwarf\_mix} and $98$ per cent for {\em Dwarf\_groups}.
\\

\section{Environmental effects} \label{S_env}

\subsection{Classification of the environment} \label{S_kfwv}

In the last few years, many papers have found evidence of 
properties of galaxies and systems of galaxies that depend on the 
environment in which they reside 
\citep[][among others]{Dressler:1980,Kauffmann:2004,Blanton:2005,
OMill:2008,Peng:2010,Wetzel2012,Zheng:2017,Wang2018,Duplancic:2020}.
There are many different methods to classify 
large-scale cosmic matter distribution \cite[see][for a review]{Libeskind2018}. 
Most of these define the Hessian matrix from the density, 
velocity or potential field using a fixed finite grid. 
Then, this matrix is diagonalised to determine their eigenvectors 
and eigenvalues, which give information about the principal directions 
and strength of local collapse or expansion. 
The main disadvantage of these methods is the finite resolution assigned by 
a finite grid. 
\cite{Wang:2020} presented an improvement to these methods given by an 
adaptive interpolation guaranteeing higher resolution. 
We apply this method to our samples and refer the reader to \cite{Wang:2020} 
for a detailed description of their method.
\\

To characterise the environment where associations 
of dwarf galaxies are found, we use the eigenvalues estimated from 
both the tidal tensor \citep[e.g.][]{Hahn:2007a} and 
shear velocity \citep{Hoffman2012} of the DM particle 
distribution of the parent simulation \smdpl.
We define four types of environment: 
knot, filament, wall or void, based on the number of 
eigenvalues larger than a chosen threshold ($\lambda_{\rm th}$). 
If we adopt the nomenclature \mbox{$\lambda_1 < \lambda_2 < \lambda_3 $} 
for the smallest, intermediate and largest, respectively, then we define

\begin{enumerate}
  \item {Knot: if $\lambda_{\rm th} < \lambda_1$ }
  \item {Filament: if $\lambda_1 < \lambda_{\rm th} < \lambda_2$ }
  \item {Wall: if $\lambda_2 < \lambda_{\rm th} < \lambda_3$ }
  \item {Void: if $\lambda_3 < \lambda_{\rm th}$}
\end{enumerate}

We use a grid of $400^3$ cells and three different specific 
smoothing lengths $l=1,\,2$ or 4 $\,{\rm Mpc}\,h^{-1}$ over 
the tidal and shear velocity tensor.
Then, we estimate the three eigenvalues for each cell. 
According to the position of the centre of mass of each association, 
we assign to it the eigenvalues of the cell where it is located.
Following \cite{Wang:2020}, we adopted a threshold 
$\lambda_{\rm th} = 0$ to define the four classifications.
In this way, each association belongs to a single 
environment: void, wall, filament or knot.
\\

The distribution of associations among the different environments 
depends on the choice of the smoothing length, as shown in 
Appendix~\ref{S_appendix}.
We have checked that the results obtained in this work do not depend 
either on the field used to classify the environment (tidal tensor 
or shear velocity), or on the smoothing length 
($l = 1,\,2$ or 4 $\,{\rm Mpc}\,h^{-1}$).
Therefore, for simplicity and clarity, the results of this paper 
will be presented only for the tidal tensor and for the smoothing 
length $ l = 1 \,{\rm Mpc}\,h^{-1}$. 
This smoothing length represents about 5 times the median size 
of the associations, so it is a sufficient volume to analyse their
environment.
\\

Note that taking $\lambda_{\rm th} = 0$, this classification takes 
into account only the sign of the eigenvalues and not the ratio 
between the eigenvalues. 
Furthermore, the tidal tensor eigenvalues refer to the directions 
of differential motion, namely compression and expansion. 
This should not be confused with the ellipsoidal shape of the 
mass distribution, defined by the inertia tensor. 
In general, however, there is a correlation and an alignment 
between the inertia tensor and the tidal tensor 
eigenvalues \citep{Porciani2002}. 
A knot means that the three eigenvalues are pointing inwards, i.e. 
the system is collapsing; a filament means that 2 eigenvalues 
are collapsing and one expanding (the expansion direction corresponds 
to the filament's `spine'). 
A wall means that 2 eigenvalues are expanding and one is collapsing (which 
corresponds to the sheet normal), and a void means that the three eigenvalues 
are pointing outwards, i.e. the system is expanding.
\\

According to this large-scale environment classification 
(knot, filament, wall, void), most of the associations 
in the {\em Dwarf\_associations} sample are located in 
filaments ($\sim 67$ per cent), followed by the wall 
environment ($\sim 26$ per cent), while a minority fraction 
is in knots ($\sim 6$ per cent) and voids ($\sim 1$ per cent). 
We specify percentage values for systems with all their galaxy 
members belonging to different main host DM haloes, i.e. 
the {\em Dwarf\_associations} sample, because they are the 
focus of our work, but these values do not change significantly 
for other samples.
Fig.~\ref{fig:histo_KFWV} shows the number 
of systems located in each large-scale environment classification, 
for the samples {\em Dwarf\_associations}, {\em Dwarf\_mix} 
and {\em Dwarf\_groups} (percentage values 
are indicated above each bar).
For the rest of the samples, these values do not change 
significantly either.
For example, for the {\em All\_associations} sample, 
$\sim 69$ per cent of the systems are located in filaments, 
$\sim 24.5$ per cent in walls, $\sim 6$ per cent in knots, and only 
$\sim 0.5$ per cent in voids. 
Likewise, if we consider systems without a restriction in the 
maximum value of the stellar mass of their 
member galaxies and without distinction of the halo in which 
their member galaxies are hosted 
(i.e. {\em All\_associations} + {\em All\_mix} + {\em All\_groups}), 
the percentages are very similar: $\sim 70$ per cent in filaments,
$\sim 22$ per cent in walls, $\sim 7.5$ per cent in knots, and 
$\sim 0.5$ per cent in voids. 
Although it is not a strictly direct comparison, 
we can compare these percentages with the mass fraction assigned to a 
given environmental classification.
As we have already mentioned above, a variety of methods have been 
developed to classify the cosmic web \cite[see][for a review]{Libeskind2018}. 
As these methods identify the web components differently, it is not 
surprising that there are significant discrepancies in these fractions. 
Despite these differences, for most of these methods, the largest mass 
fraction is found in filaments (ranging from $\sim30$ to $\sim90$ per cent 
depending on the method), followed by walls or nodes depending on the method, 
while the lowest mass fraction is found in voids (less than 10 per cent in most 
of these methods). 
Despite not being a direct comparison and taking into account the differences 
in the percentages, our results follow the same trend presented by these works.
\\

\begin{figure}
    \begin{subfigure}{
        \includegraphics[width=\columnwidth]{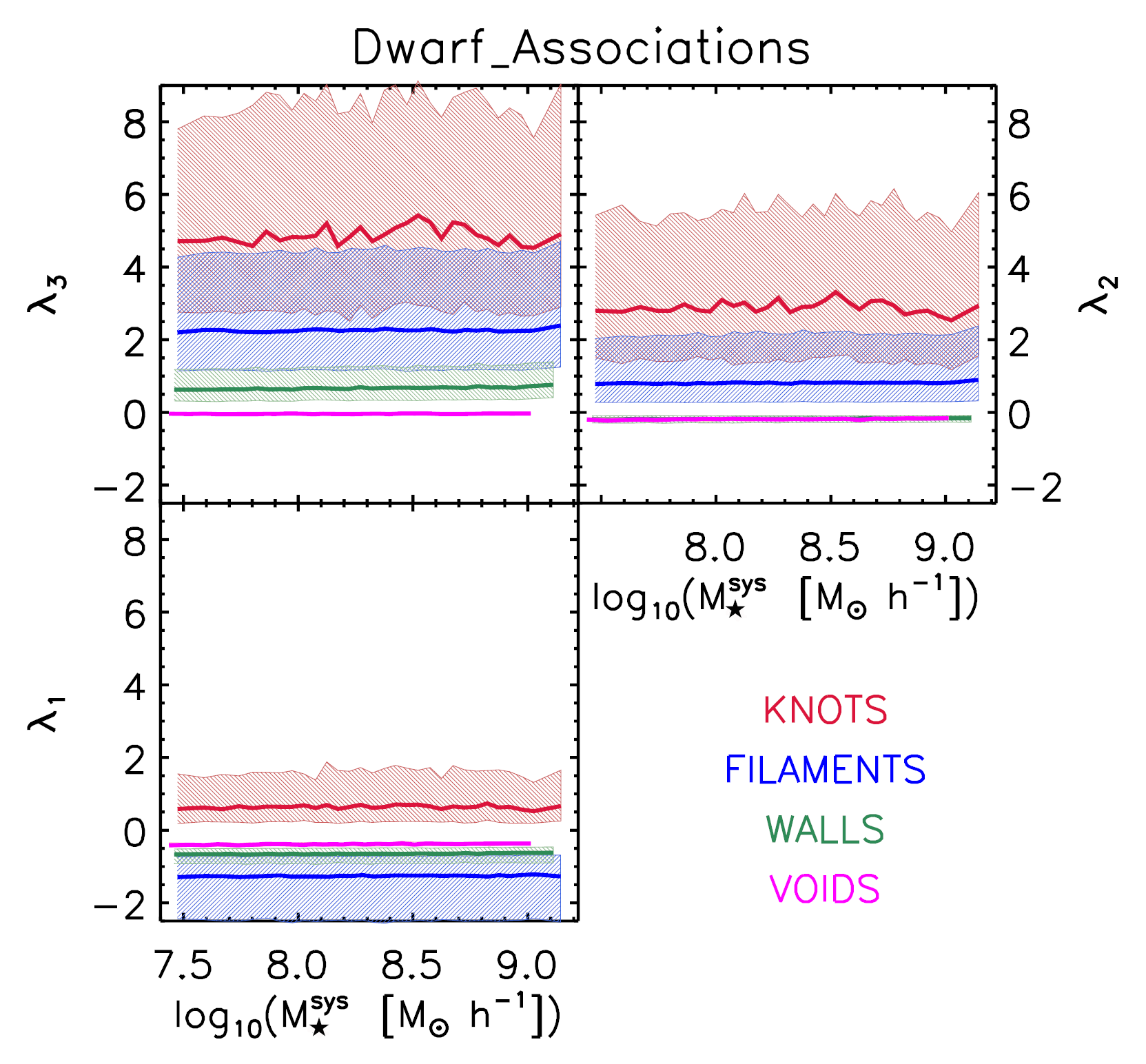}}
    \end{subfigure}
    \caption{Values of the three eigenvalues as a function of 
    the stellar mass of associations for the sample {\em Dwarf\_associations} 
    split according to the environment in which they reside
    as indicated by the legends. 
    Coloured solid lines show median values of 
    eigenvalues taking equal number of bins in $M_{\star}^{\rm sys}$.
    Shaded regions cover from $25$ per cent 
    to $75$ per cent for each sample.}
    \label{fig:eigenvalues}
\end{figure}

Fig.~\ref{fig:eigenvalues} shows values of the three eigenvalues 
as a function of the total stellar mass of the systems 
($M_{\star}^{\rm sys}$), for the sample 
{\em Dwarf\_associations}, split according to the environment in 
which they reside as indicated by the legends. 
Coloured lines show median values of eigenvalues taking 
equal number of bins in $M_{\star}^{\rm sys}$, 
while shaded regions cover from $25$ per cent to $75$ per cent 
for each sample.
The eigenvalues do not show a significant dependence on the stellar 
mass of the association. 
From this figure, the sign of each eigenvalue is evident,
indicating the direction of the tidal field (collapse if positive 
and expansion if negative) taking into account the adopted threshold 
$\lambda_{\rm th} = 0$ for the classification of the structure in the cosmic web.

\begin{figure*}
    \begin{subfigure}{
        \includegraphics[width=0.5\textwidth]{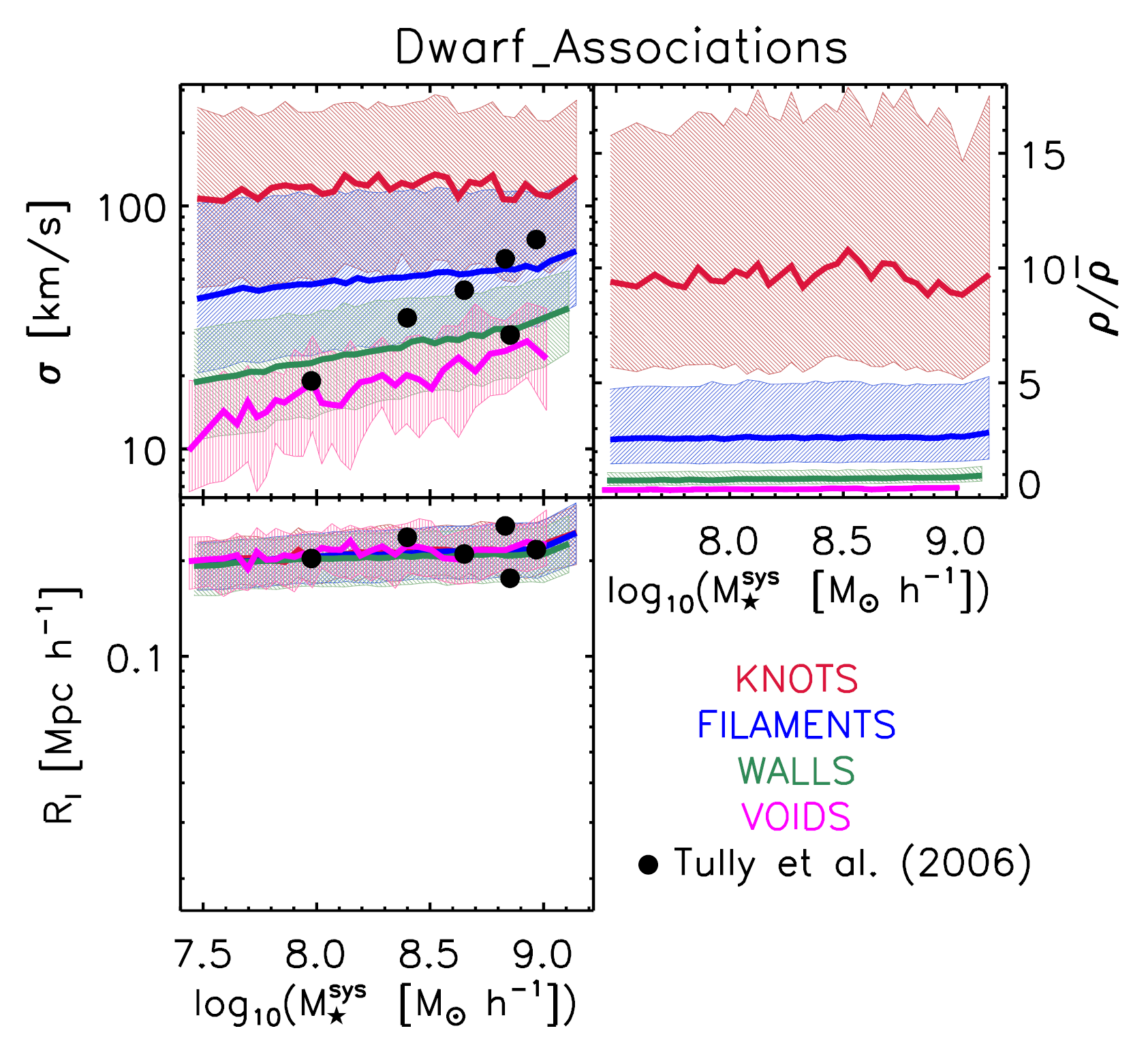}
        \includegraphics[width=0.5\textwidth]{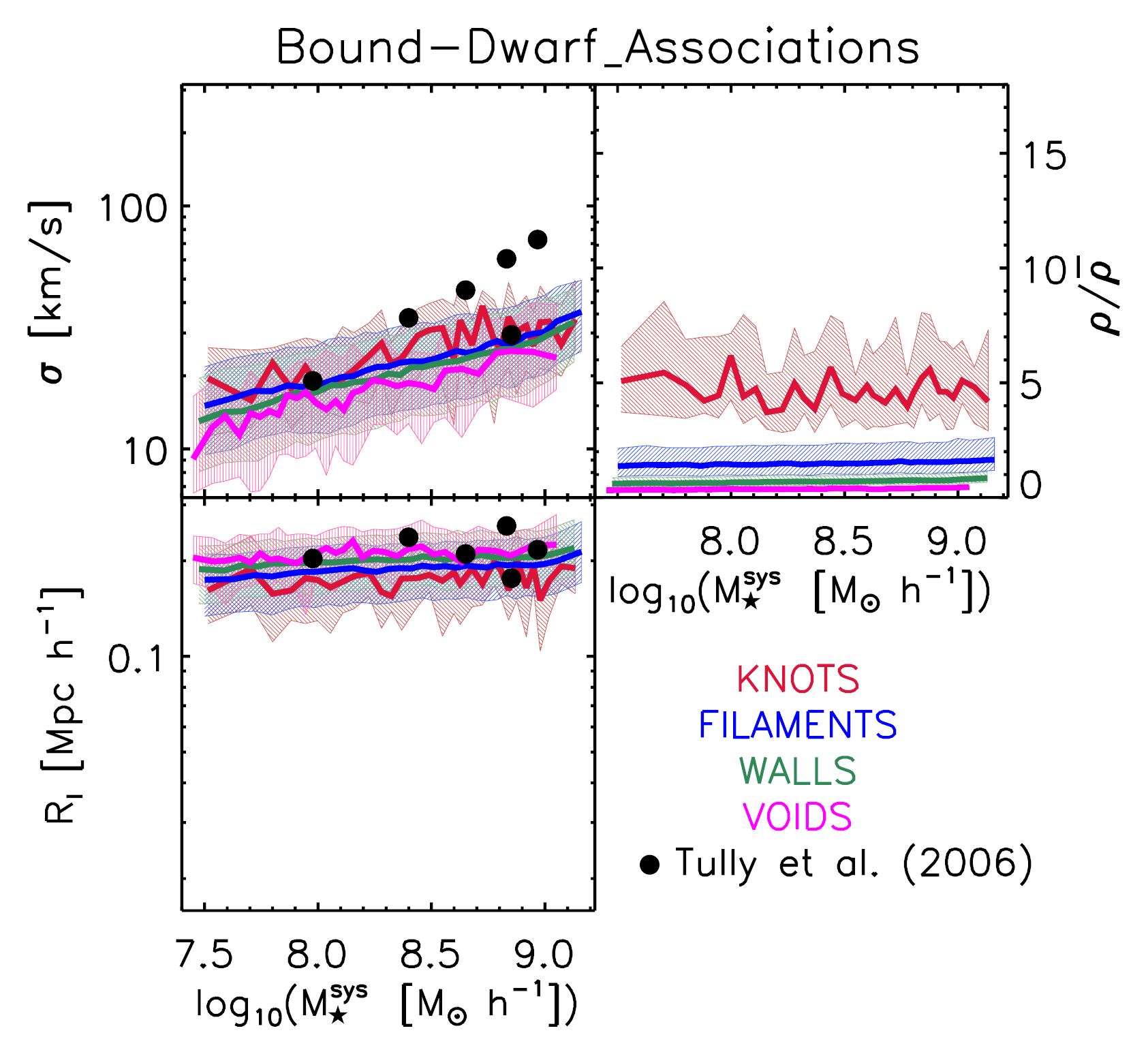}}
    \end{subfigure}
    \caption{Left panels: Dynamical properties of associations 
    as a function of the stellar mass of the association ($M_{\star}^{\rm sys}$) 
    for systems in the sample {\em Dwarf\_associations}: 
    velocity dispersion ($\sigma$, top left),  
    size ($R_{\rm I}$, bottom left) and the ratio between the DM particles
    density $\rho$ in the grid cell where the system is located and the mean density 
    ($\rho/\bar{\rho}$, top right). 
    Solid lines show median values of each property, 
    taking equal number of bins in $M_{\star}^{\rm sys}$. 
    Each colour corresponds to a different environment (knots, filaments, walls and voids) 
    as indicated in the legends. 
    Shaded regions cover from $25$ per cent to $75$ 
    per cent for each sample.
    Results for simulated associations of dwarf galaxies are compared with 
    observational results taken from \citet[][black filled circles]{Tully:2006}, 
    assuming for the latter a factor equal to one in the mass-to-light ratio.
    Rigth panels: Idem as left panels but considering a sub--sample 
    with only gravitationally bound systems ({\em Bound--Dwarf\_associations}).}
    \label{fig:prop_diff}
\end{figure*}

\subsection{Properties of dwarf galaxy associations} \label{S_prop}

To analyse the environmental effects on the associations 
of dwarf galaxies, we study how their main dynamical 
properties vary according to the environment in which 
these particular systems reside. 
As the main intrinsic properties of our systems, we compute the
inertial radius ($R_{\rm I}$, defined in eq.~(\ref{eq:radius})) as an indicator 
of the size of the system, the velocity dispersion ($\sigma$), and 
the stellar mass of the system ($M_{\star}^{\rm sys}$).
The velocity dispersion is computed through the unbiased 
sample variance of the sample for all member galaxies, defined by  

\begin{equation}
\sigma =\left [\sum_{i}^{N} v_{i}^{2}/(N-1)\right]^{1/2}.
\label{eq:vel}
\end{equation}

\noindent where $v_i$ is the one-dimensional velocity 
difference between a galaxy and the mean velocity of the system.
\\

Fig.~\ref{fig:prop_diff} shows the main dynamical properties of our systems 
($R_{\rm I}$ and $\sigma$) as a function of the total stellar 
mass of the association ($M_{\star}^{\rm sys}$)
for systems in the sample of dwarf galaxy associations
({\em Dwarf\_associations}, left panels). 
Coloured lines show median values of each property taking 
an equal number of bins in $M_{\star}^{\rm sys}$.
Each colour corresponds to a different environment 
(knots, filament, walls and voids), as indicated by the legends. 
Top right panels show the ratio between the DM particles 
density $\rho$ in the grid cell where the system is located and the mean density 
($\rho/\overline{\rho}$) as a function of the stellar mass 
of the system.
This ratio is related with eigenvalues by
\mbox{$\rho/\overline{\rho} - 1 = \lambda_{1} + \lambda_{2} + \lambda_{3}$}. 
As expected, it is clear that knots correspond to the highest 
densities, followed by filaments, then walls and finally voids, 
which correspond to the lowest density.
From the results for the associations 
({\em Dwarf\_associations} sample),
it is evident that the velocity dispersion depends strongly on the 
environment where the association is located. 
Associations have a very low velocity dispersion 
($\sigma \sim 20 \,{\rm km\,s^{-1}}$) 
if found in voids, which increases up to $\sigma \sim 140 \, {\rm km\,s^{-1}}$ 
as we move to knots, going through walls and filaments.
As associations are composed exclusively of 
galaxies that do not live in the same DM halo, they are somehow 
tracing the velocity dispersion of the environment where they are located: 
knot, filament, wall or void.
This dependence of velocity dispersion on the environment has 
already been previously studied in some published works.
Among the most recent, we can mention \cite{Taverna:2023} who study the
effects of different global environments on the properties of Hickson-like 
compact groups of galaxies (CGs) identified in the Sloan Digital Sky Survey 
Data Release 16 \citep{Ahumada:2020}. 
They found that CG velocity dispersion increases with
the density of the environment they inhabit, since the median velocity dispersion 
observed for CGs in the highest-density environments almost
doubles that observed for CGs in the lowest-density environments.
Furthermore, we can mention the results presented by \cite{Ruiz:2019} who
compare pairwise velocity ($w$) distributions for all galaxies with pairwise 
velocity distributions for galaxies located in void regions.
They found $ w \sim 500 \,{\rm km\,s^{-1}}$ for all galaxies, while 
for void galaxies the pairwise velocity dispersions are in the range 
$w \sim 50-70 \,{\rm km\,s^{-1}}$, roughly one order of magnitude smaller. 
They found these differences in both the observations and in the simulated
galaxies.
Our results are in accordance with these results in the sense that 
the velocity dispersion shows the same trend, being significantly 
smaller in low-density regions than in high-density regions.
Moreover, this progressive increase of the values 
of the velocity dispersion through different environments, 
regardless of the total stellar mass of the association, 
is also accompanied by an increasing trend of the velocity dispersion 
with increasing $M_{\star}^{\rm sys}$, and this trend is more pronounced 
for associations residing in filaments and almost non-existent 
in associations located in knots. 
In relation to the size of the associations, 
there are no systematic effects of the environment on 
the size of associations.
\\
\\
Compared with observational results, the dynamical properties 
of most of the observed dwarf galaxy associations presented 
by \cite{Tully:2006} (black filled circles) are compatible 
at first glance with a filament-like (3 systems) or wall-like (2 systems)
environment, while just 1 association is compatible with 
a void-like environment.
In a deeper analysis, we compute the probability
of each association to belong to a web-type environment classification. 
We estimate the probability distribution function (PDF) of velocity dispersion 
for each web-type environment classification, regarding bins of mass.
Taking this distribution into account we assign a probability value for 
each observed association.
This probability is shown in Table~\ref{table:prob}, where the highest 
probability for each association is highlighted in bold type. 
The associations are stored in increasing order of their velocity dispersion.
Considering the highest values of these probabilities, it appears that 
two associations belong to the void-like environment 
(Association 1 and Association 2), two belong to the wall-like environment 
(Association 3 and Association 4), and two belong to the filament-like 
environment (Association 5 and Association 6).
Nevertheless, these highest probabilities are not significantly 
larger compared to the others, thereby indicating that the membership 
in a given web-type classification is not particularly evident.
This is primarily due to the large scatter in velocity dispersion 
for each environment, as evidenced by the shaded regions in the figure.
Another important thing to note is that, as the velocity dispersion of the 
associations increases, the probability of belonging to denser environments 
also increases.
\\

We also analyze systems in the 
sample built without restriction in the maximum value of the 
galaxy stellar mass ({\em All\_associations}) and there are 
not significant differences between their results 
and those previously described.
Due to the similarities between results, we do not show the 
latter ones to make the discussion clearer.
In brief, these results show that the increase of $\sigma$ with the 
density of the environment does not happen only in associations 
of dwarf galaxies, but is noticed in all systems where their
member galaxies belong to different DM haloes.
\\
\\

\begin{table}
\centering
   \begin{tabular}{  c  c  c  c  c  c  c  }
    \hline
    \hline
    \\
       \multicolumn{1}{c}{Observed} & \multicolumn{1}{c}{$\sigma [{\rm km/s}]$} & 
       \multicolumn{1}{c}{Knot} & \multicolumn{1}{c}{Filament} & 
       \multicolumn{1}{c}{Wall} & \multicolumn{1}{c}{Void} & 
    \\
       \multicolumn{1}{c}{Associations} & & & & & &
    \\  
    \hline
     1 & 19.05 & 6$\%$ & 16$\%$ & 34$\%$ & \textbf{44$\%$} & 
    \\
     2 & 29.44 & 6$\%$ & 21$\%$ & 33$\%$ & \textbf{40$\%$} &
    \\ 
     3 & 34.64 & 10$\%$ & 26$\%$ & \textbf{36$\%$} & 28$\%$ &
    \\
     4 & 45.03 & 17$\%$ & 30$\%$ & \textbf{32$\%$} & 20$\%$ &
    \\
     5 & 60.62 & 24$\%$ & \textbf{35$\%$} & 29$\%$ & 12$\%$ &
    \\
     6 & 72.75 & 34$\%$ & \textbf{39$\%$} & 27$\%$ & 0$\%$ &
    \\
    \hline
    \hline
  \end{tabular}
  \caption{Probability of each observed association 
  to belong to one environment. 
  First column indicates the association. 
  Second column shows the velocity dispersion of the association.
  Third, fourth, fifth and sixth columns indicate the probability of each 
  observed association to belong to one web type environment: 
  knot, filament, wall and void, respectively. 
  The highest probability for each association is highlighted in bold type.}
 \label{table:prob}
\end{table}

Going even further, right panels of 
Fig.~\ref{fig:prop_diff} show the same as left panels but 
considering only gravitationally bound systems 
({\em Bound--Dwarf\_associations}).
For this sample, it is evident that the dependence of velocity 
dispersion on the environment where the association is located 
is noticeably attenuated. 
The velocity dispersion covers a narrow range around low 
velocity dispersions, between 
$\sigma \sim 10 \, {\rm km\,s^{-1}}$ and 
$\sigma \sim 30 \, {\rm km\,s^{-1}}$, depending on the mass of the 
system, with minimal differences according to the environment in 
which they are located.
Although a trend is slightly visible, it is too weak.
Therefore, we can infer that the velocity dispersion of 
gravitationally bound systems does not depend on the environment 
in which they reside.
Comparing with observational results, the velocity dispersion of 
at least half of the observed dwarf galaxy associations presented 
by \cite{Tully:2006} (black filled circles) are not compatible 
with our results.
This would indicate that these observed associations could be 
non--gravitationally bound systems. 
\\

\begin{figure*}
    \begin{subfigure}{
        \includegraphics[width=0.5\textwidth]{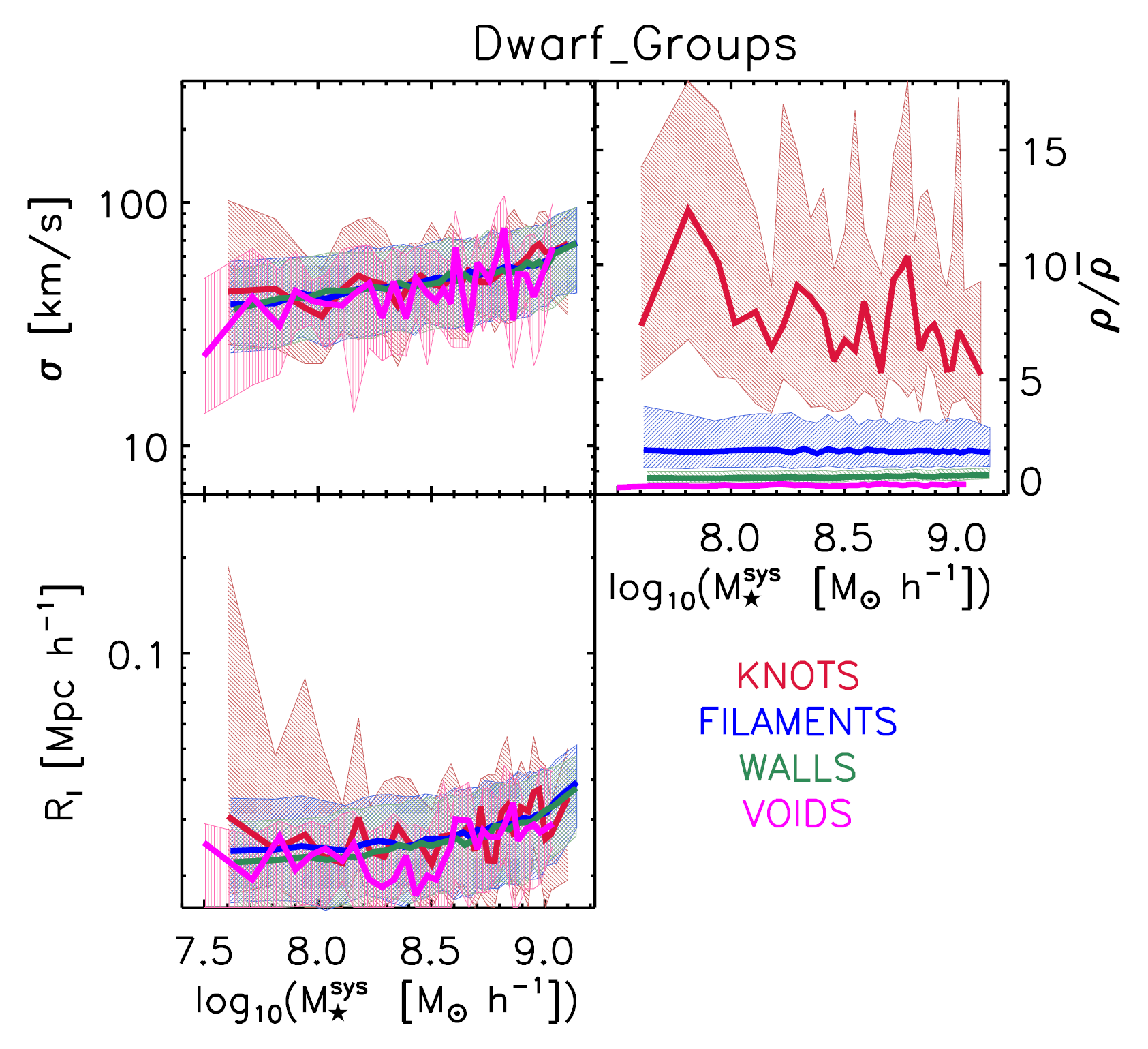}
        \includegraphics[width=0.5\textwidth]{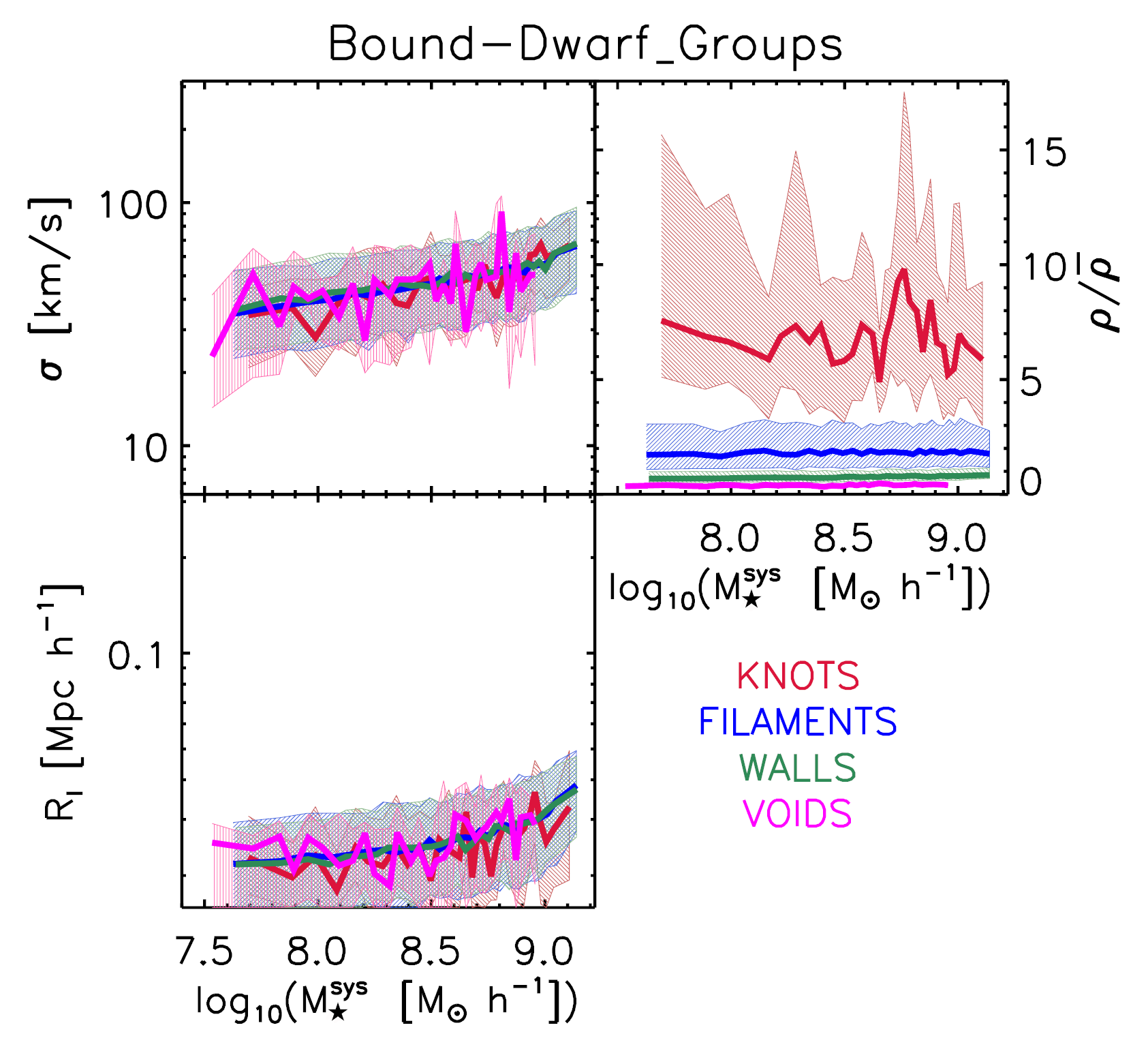}}
    \end{subfigure}
    \caption{
    Left panels: Dynamical properties of groups of galaxies as a function of the 
    stellar mass of the group ($M_{\star}^{\rm sys}$) for systems in the sample 
    {\em Dwarf\_groups}: velocity dispersion ($\sigma$, top left), 
    size ($R_{\rm I}$, bottom left) and the ratio between the DM particles 
    density $\rho$ in the grid cell where the system is located and the mean density 
    ($\rho/\bar{\rho}$, top right). 
    Solid lines show median values of each property, 
    taking an equal number of bins in $M_{\star}^{\rm sys}$. 
    Each colour corresponds to a different environment (knots, filaments, 
    walls and voids) as indicated in the legends.
    Shaded regions cover from $25$ per cent to $75$ 
    per cent for each sample.
    Rigth panels: Idem as left panels but considering a sub--sample 
    with only gravitationally bound systems: {\em Bound--Dwarf\_groups}.}
    \label{fig:prop_same}
\end{figure*}

Fig.~\ref{fig:prop_same} shows the same relationships as 
Fig.~\ref{fig:prop_diff} but considering different samples.
Panels in the left box correspond to systems of sample 
{\em Dwarf\_groups} while panels in the right box correspond 
to systems of sample 
{\em Bound--Dwarf\_groups}, a sub--sample 
of the former containing only gravitationally bound systems.
In these cases, unlike the samples shown in the previous figure, 
all member galaxies belong to the same main host DM halo.
It is noticeable that the typical size of these systems is much 
smaller than those of samples {\em Dwarf\_associations} 
and {\em Bound--Dwarf\_associations} shown in Fig.~\ref{fig:prop_diff}.
Moreover, the dependence of $\sigma$ on the environment
in which the systems reside disappears. 
It is evident that the behaviour of $\sigma$ with respect to the 
stellar mass of the system is very similar for all environments. 
As expected, the results of these two samples 
are very similar between them, due to the fact that 
{\em Bound--Dwarf\_associations} is a sub--sample of 
{\em Dwarf\_groups}, which contains about $\sim 96$ per cent of 
their groups. 
The remaining $\sim 4$ per cent, that is, the unbound groups of dwarf 
galaxies, mainly consist of subgroups found within larger DM halos.
We also analyze systems in the 
sample built without an upper limit in their stellar mass 
({\em All\_groups}) and there are not significant differences 
between their results and those just described. 
Again, we avoid showing the results of this sample 
for a less confusing discussion.
So, in summary, velocity dispersion does not depend 
on the environment when all galaxy members of the system belong 
to the same main DM halo.  
\\

Comparing results from Figs.~\ref{fig:prop_diff} 
and~\ref{fig:prop_same}, we infer that, when estimating the
velocity dispersion $\sigma$ for systems that belong to 
the {\em Dwarf\_associations} sample, we are actually 
estimating the velocity dispersion of the environment 
in which the associations are immersed, immediately 
surrounding them, and not of the system itself. 
Since galaxy members of these systems belong to different 
main DM haloes, they are located far from each other.
When estimating velocity dispersions of a set of galaxies 
that are far away from each other, the properties of the 
environment inevitably affect this estimation.
This occurs not only for associations of dwarf galaxies, 
but for all systems where all their member galaxies belong 
to different main DM haloes ({\em Dwarf\_associations} and 
{\em All\_associations} samples).
In contrast, when we consider gravitationally bound systems, 
the difference in the velocity dispersion with the environment
is noticeably attenuated, completely disappearing if we consider 
systems where all their member galaxies belong to the 
same main DM halo. 
This is true for systems built up only of dwarf galaxies 
and for systems without restriction in the stellar mass 
of their member galaxies, 
i.e., {\em Bound--Dwarf\_associations}, 
{\em Bound--Dwarf\_groups}, {\em Dwarf\_groups} and {\em All\_groups} samples.
In summary, considering systems whose galaxies 
belong to the same main DM halo or gravitationally bound systems 
(despite the fact that their member galaxies may belong to 
different main DM haloes), the environment does not play a fundamental 
role when we estimate the velocity dispersion of their member galaxies.
The size of dwarf galaxy systems are directly related to the DM halo 
in which galaxies reside, while the velocity dispersion of dwarf galaxy 
systems are directly related to the binding energy of the system and not to the 
environment in which the DM halo is found within the cosmic web.
\\

\begin{figure*}
    \begin{subfigure}{
        \includegraphics[width=0.5\textwidth]{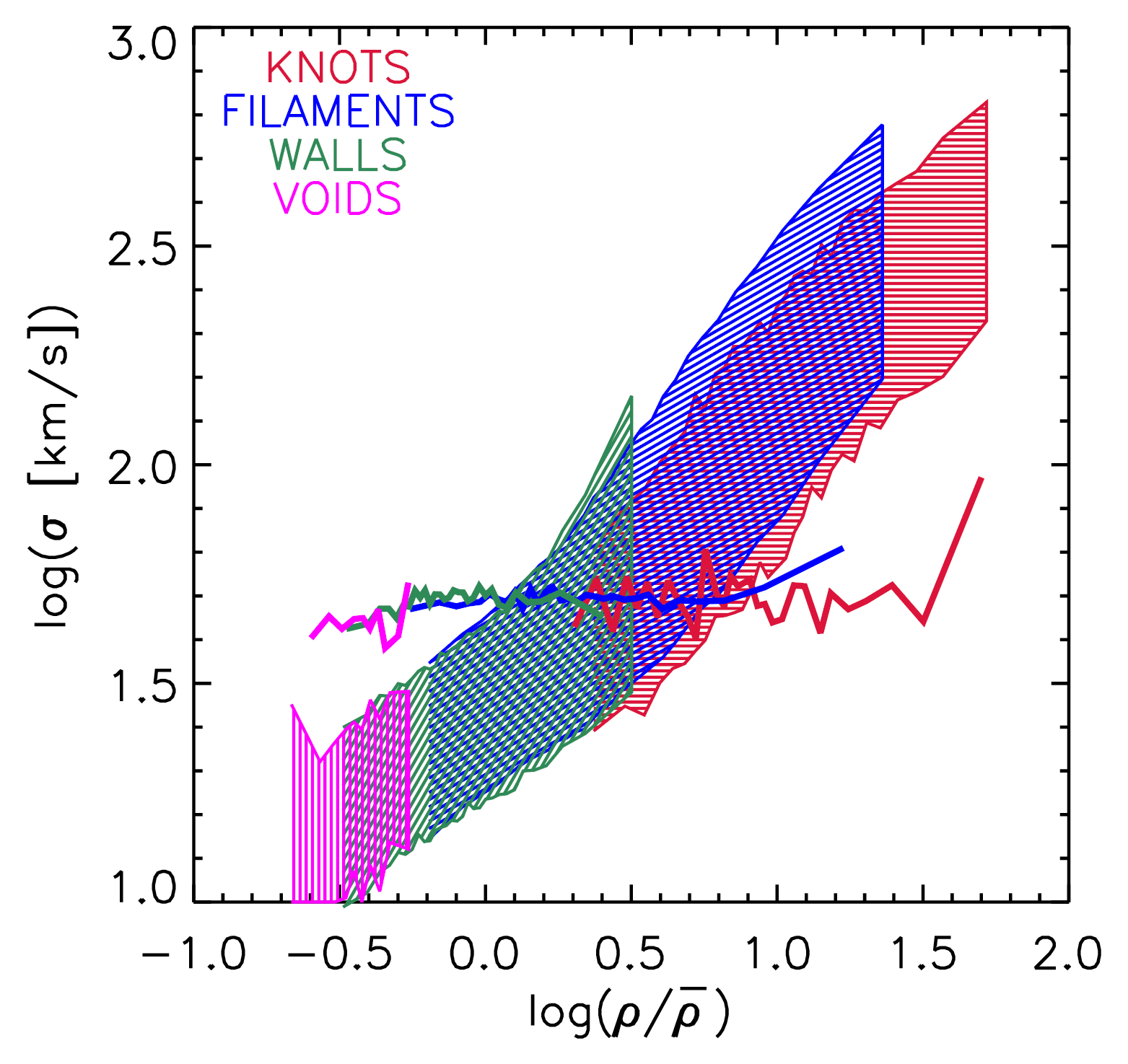}
        \includegraphics[width=0.5\textwidth]{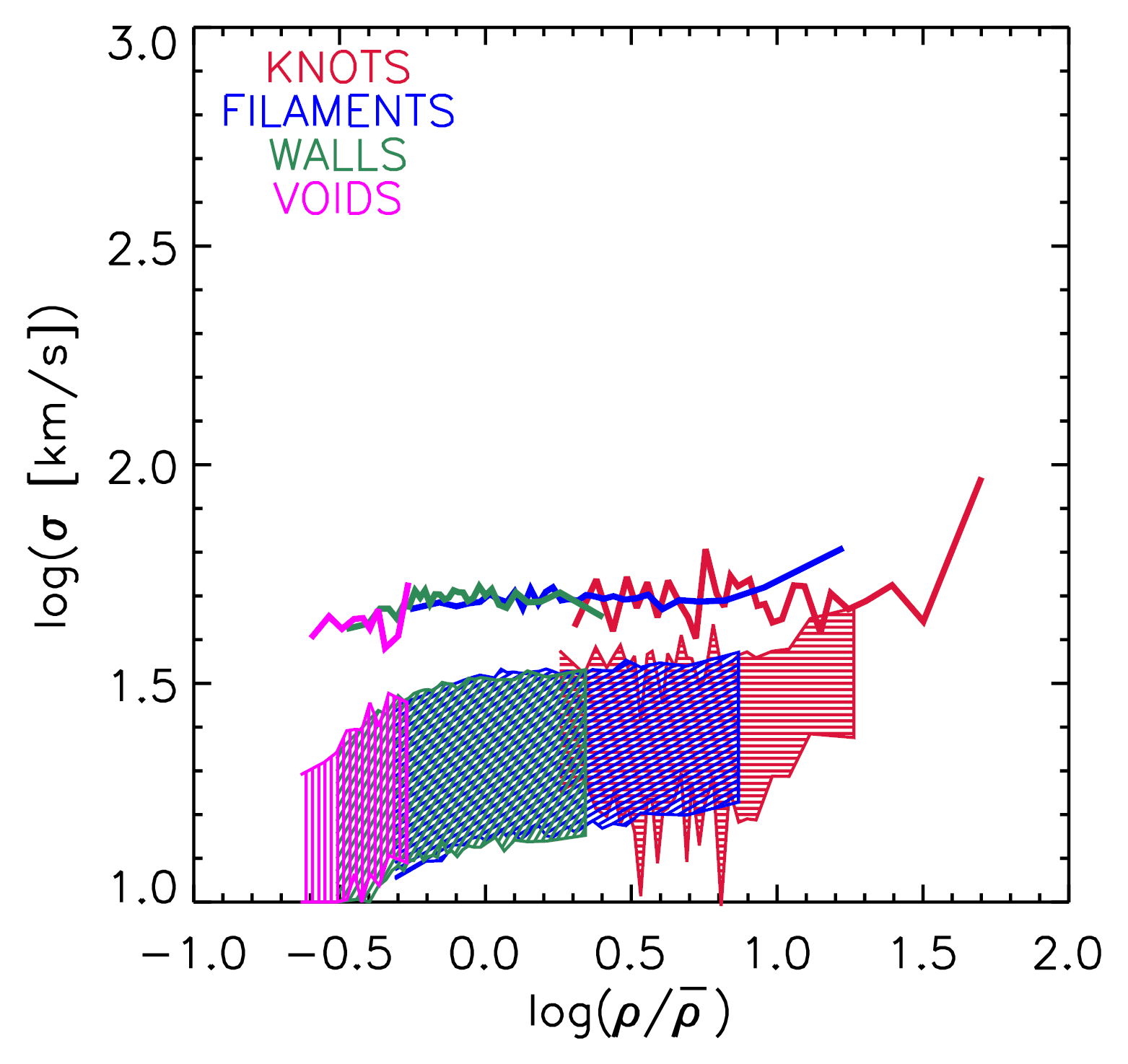}}
    \end{subfigure}
    \caption{Left panel: Velocity dispersion ($\sigma$) as a function of DM particle 
    density $\rho$ in the grid cell where the system is located divided by the mean density
    ($\rho/\bar{\rho}$), shown for two different samples, {\em Dwarf\_associations} 
    (shaded regions) and {\em Dwarf\_groups} (solid lines), for four different environments, 
    as indicated by the colour code. 
    Shaded regions cover from $25$ per cent to $75$ per cent for the {\em Dwarf\_associations} sample.
    For the case of {\em Dwarf\_groups}, we show only the median values 
    (i.e. corresponding to $50$ per cent of the sample) to make the plot clearer. 
    Rigth panel: Idem as left panel but considering sub--samples with only 
    gravitationally bound systems: {\em Bound--Dwarf\_associations} and {\em Bound--Dwarf\_groups}.}
    \label{fig:sigma_env}
\end{figure*}

The results that we have just reached can be better 
visualized by Fig.~\ref{fig:sigma_env}. 
Here, the velocity dispersion ($\sigma$) of the galaxy system
as a function of the density field of the environment 
normalised by the mean density ($\rho/\bar{\rho}$) is shown 
for different samples.
Left panel shows {\em Dwarf\_associations} 
(shaded regions) and {\em Dwarf\_groups} (solid lines) samples,
residing in different environments (indicated by the colour code).
Shaded regions cover from $25$ per cent to $75$ per cent 
of each sample of associations of dwarf galaxies; for dwarf groups, 
we show only median values (i.e., corresponding to 
$50$ per cent of the sample) to make the figure clearer.
In the case of {\em Dwarf\_associations}, the dependence of 
the velocity dispersion on density is remarkable. 
As the density of the system environment increases, the velocity 
dispersion of the system also increases.
On the other hand, it is clear that the velocity dispersion of 
{\em Dwarf\_groups} does not depend on the density of the 
environment in which the system resides.
As the density of the system environment increases, the velocity 
dispersion does not show significant changes.
This shows that the velocity dispersion of a system where 
all its member galaxies belong to different main DM haloes 
depends strongly on the environment in which the system is located.
In contrast, the velocity dispersion of a system where all  
its member galaxies belong to the same main DM halo does not 
depend on the environment of the system.
On the other side, right panels show 
samples where all systems are gravitationally bounded: 
{\em Bound--Dwarf\_associations} (shaded regions) and 
{\em Bound--Dwarf\_groups} (solid lines) samples. 
This comparison shows that velocity dispersion does not 
significantly depend on the environment when we restrict our 
samples to gravitationally bound systems. 
This figure confirms that the velocity dispersion of
dwarf galaxy systems are directly related to the binding energy of
the system and not to the environment in which the DM halo is
found within the cosmic web, regardless whether member galaxies are 
hosted by the same main DM halo or by different main DM haloes.
\\

\section{Conclusions} \label{S_conclusions}

We analyze how the dynamical properties of associations of 
dwarf galaxies depend on their environment. 
Within the \mbox{$\Lambda$CDM} cosmological context, we identify 
these particular systems in the high-resolution dark matter-only 
\smdpl~simulation \citep{Klypin:2016}, coupled to the 
\sag~semi-analytical model of galaxy formation \citep{Cora:2018}. 
We compare associations of dwarf galaxies, where all
their members belong to different main DM haloes, with close 
groups of dwarf galaxies hosted by a single massive DM halo.
From \cite{Yaryura:2020} we know that the so defined associations 
of dwarf galaxies are the systems which best reproduce the dynamical 
properties of observed associations presented by \cite{Tully:2006}.  
That is why studying associations of dwarf galaxies is our main objective.  
\\

We classify the environment into the four different categories of knots, 
filaments, walls and voids, and analyse its effect on the main properties 
of the associations and groups of dwarf galaxies. 
Most dwarf galaxies associations are located in filaments ($\sim 67$ per cent), 
followed by walls ($\sim 26$ per cent), knots ($\sim 6$ per cent) and 
voids ($\sim 1$ per cent). 
So far, only seven associations of dwarf galaxies have been observed \citep{Tully:2006}.  
Based on the observed velocity dispersion we conclude that three of them 
are most compatible with a filament-like environment, two with a 
wall-like environment, while just one is most compatible with a void-like environment. 
Based on the PDF of the velocity dispersion of associations in 
different environments we estimated the probabilities of the observed associations 
to be located in a specific environment (Table~\ref{table:prob}) and found that 
they tend to be located in low density environment (voids, walls up to filaments) 
and most probably cannot be found in knots. 
It is worth noting that as the velocity dispersion of the associations increases, 
so does the probability of belonging to denser environments.
\\

When we analyze the dependence of the dynamical properties of the associations 
of dwarf galaxies on the environment, we find that the velocity dispersion 
strongly depends on the environment.  
Associations have a very high-velocity dispersion 
($\sigma \sim 140 \; {\rm km\,s^{-1}}$) if located in knots, 
which decreases to $\sigma \sim 20 \; {\rm km\,s^{-1}}$ as we reach voids, 
going through filaments and walls.  
This dependency occurs in systems composed exclusively of dwarf galaxies 
as well as in systems with no restriction on the maximum value of the 
stellar mass of their member galaxies.  
Since the members of the associations belong to different DM haloes, 
these galaxy systems could be considered as large-scale mass distribution tracers.  
The velocity dispersion of these galaxies reflects the velocity dispersion 
of the environment (knot, filament, wall or void) in which they reside.  
Therefore large-scale environment plays a fundamental role in determining 
the dynamical properties of associations.
\\

Being more restrictive in the definition of associations, requiring them 
to be gravitationally bound systems, our results change significantly.  
When we consider a sub--sample of associations of dwarf galaxies, 
with only gravitationally bound systems ($\sim30$ per cent of the total 
sample of associations of dwarf galaxies), the dependence of 
the velocity dispersion on the environment is strongly attenuated.  
This indicates that the environment significantly influences the 
dynamical properties of systems only when they are not physically bound systems.  
Comparing with observational results presented by \cite{Tully:2006}, 
the velocity dispersion of most of these associations is not compatible 
with that of the bound systems, which could indicate that these 
observed associations would not be gravitationally bound systems.
\\

When we focus on the groups of dwarf galaxies, defined as 
systems where all member galaxies belong to the same main DM halo, 
we do not find any dependence of the velocity dispersion on the environment.
About $\sim 96$ per cent of the groups are gravitational bound systems. 
So, the dynamical properties of these groups are not influenced by 
the environment where they reside.
\\

\section*{Acknowledgements}

We would like to thank the referee for carefully reading the 
manuscript and making a lot of suggestions which improved our paper 
substantially.
The SMDPL simulation was performed at LRZ Munich within the pr87yi project. 
The authors gratefully acknowledge funding for this project from the Gauss Centre for Supercomputing e.V.
(www.gauss-centre.eu) by providing computing time 
on the GCS Supercomputer SUPERMUC-NG at the Leibniz Supercomputing Centre 
(www.lrz.de).
The CosmoSim database (www.cosmosim.org) used in this paper is a service of the 
Leibniz Institute for Astrophysics Potsdam (AIP).
Our collaboration has been supported by the DFG grant GO 563/24-1.
This work has been partially supported by the Consejo de
Investigaciones Cient\'ificas y T\'ecnicas de la Rep\'ublica Argentina
(CONICET) 
and the Agencia Nacional de 
Promoci\'on Cient\'ifica y Tecnol\'ogica (PICT 2019-1600).
NIL acknowledges financial support from the IDEXLYON Project at the University of Lyon under the Investments for the Future Program (ANR-16-IDEX-0005). NIL also acknowledge support from the joint Sino-German DFG research Project ``The Cosmic Web and its impact on galaxy formation and alignment'' (DFG-LI 2015/5-1).
SAC acknowledges funding from CONICET (PIP-2876), {\it 
Agencia Nacional de Promoci\'on de la Investigaci\'on, el Desarrollo Tecnol\'ogico y la Innovaci\'on} (Agencia I+D+i, PICT-2018-3743), and the {\it Universidad Nacional de La Plata} (G11-150), Argentina.
ANR acknowledges support from CONICET (PIP 11220200102832CO) and the Secretar\'ia de Ciencia y T\'ecnica de la Universidad Nacional de C\'ordoba (PID 33620180101077).
CVM acknowledges support from ANID/FONDECYT through grant 3200918, and he
also acknowledges support from the Max Planck Society through
a Partner Group grant.
GY acknowledges partial financial support from the Ministerio de  Ciencia e Innovación (Spain) under research grant PID2021-122603NB-C21.

\section*{Data availability}

The data underlying this article will be shared on reasonable request to the corresponding author.



\bibliographystyle{mnras}
\bibliography{Bibliography.bib} 

\begin{thebibliography}{}
\makeatletter
\relax
\def\mn@urlcharsother{\let\do\@makeother \do\$\do\&\do\#\do\^\do\_\do\%\do\~}
\def\mn@doi{\begingroup\mn@urlcharsother \@ifnextchar [ {\mn@doi@}
  {\mn@doi@[]}}
\def\mn@doi@[#1]#2{\def\@tempa{#1}\ifx\@tempa\@empty \href
  {http://dx.doi.org/#2} {doi:#2}\else \href {http://dx.doi.org/#2} {#1}\fi
  \endgroup}
\def\mn@eprint#1#2{\mn@eprint@#1:#2::\@nil}
\def\mn@eprint@arXiv#1{\href {http://arxiv.org/abs/#1} {{\tt arXiv:#1}}}
\def\mn@eprint@dblp#1{\href {http://dblp.uni-trier.de/rec/bibtex/#1.xml}
  {dblp:#1}}
\def\mn@eprint@#1:#2:#3:#4\@nil{\def\@tempa {#1}\def\@tempb {#2}\def\@tempc
  {#3}\ifx \@tempc \@empty \let \@tempc \@tempb \let \@tempb \@tempa \fi \ifx
  \@tempb \@empty \def\@tempb {arXiv}\fi \@ifundefined
  {mn@eprint@\@tempb}{\@tempb:\@tempc}{\expandafter \expandafter \csname
  mn@eprint@\@tempb\endcsname \expandafter{\@tempc}}}

\bibitem[\protect\citeauthoryear{{Ahumada} et~al.,}{{Ahumada}
  et~al.}{2020}]{Ahumada:2020}
{Ahumada} R.,  et~al., 2020, \mn@doi [\apjs] {10.3847/1538-4365/ab929e}, \href
  {https://ui.adsabs.harvard.edu/abs/2020ApJS..249....3A} {249, 3}

\bibitem[\protect\citeauthoryear{{Arag{\'o}n-Calvo}, {van de Weygaert}, {Jones}
   \& {van der Hulst}}{{Arag{\'o}n-Calvo} et~al.}{2007}]{Aragon:2007}
{Arag{\'o}n-Calvo} M.~A.,  {van de Weygaert} R.,  {Jones} B. J.~T.,   {van der
  Hulst} J.~M.,  2007, \mn@doi [\apjl] {10.1086/511633}, \href
  {https://ui.adsabs.harvard.edu/abs/2007ApJ...655L...5A} {655, L5}

\bibitem[\protect\citeauthoryear{{Behroozi}, {Wechsler}  \& {Wu}}{{Behroozi}
  et~al.}{2013a}]{Behroozi_rockstar}
{Behroozi} P.~S.,  {Wechsler} R.~H.,   {Wu} H.-Y.,  2013a, \mn@doi [\apj]
  {10.1088/0004-637X/762/2/109}, \href
  {http://cdsads.u-strasbg.fr/abs/2013ApJ...762..109B} {762, 109}

\bibitem[\protect\citeauthoryear{{Behroozi}, {Wechsler}, {Wu}, {Busha},
  {Klypin}  \& {Primack}}{{Behroozi} et~al.}{2013b}]{Behroozi_ctrees}
{Behroozi} P.~S.,  {Wechsler} R.~H.,  {Wu} H.-Y.,  {Busha} M.~T.,  {Klypin}
  A.~A.,   {Primack} J.~R.,  2013b, \mn@doi [\apj]
  {10.1088/0004-637X/763/1/18}, \href
  {http://adsabs.harvard.edu/abs/2013ApJ...763...18B} {763, 18}

\bibitem[\protect\citeauthoryear{{Blanton}, {Eisenstein}, {Hogg}, {Schlegel}
  \& {Brinkmann}}{{Blanton} et~al.}{2005}]{Blanton:2005}
{Blanton} M.~R.,  {Eisenstein} D.,  {Hogg} D.~W.,  {Schlegel} D.~J.,
  {Brinkmann} J.,  2005, \mn@doi [\apj] {10.1086/422897}, \href
  {http://adsabs.harvard.edu/abs/2005ApJ...629..143B} {629, 143}

\bibitem[\protect\citeauthoryear{{Boselli}, {Cortese}, {Boquien}, {Boissier},
  {Catinella}, {Lagos}  \& {Saintonge}}{{Boselli} et~al.}{2014}]{boselli14}
{Boselli} A.,  {Cortese} L.,  {Boquien} M.,  {Boissier} S.,  {Catinella} B.,
  {Lagos} C.,   {Saintonge} A.,  2014, \mn@doi [\aap]
  {10.1051/0004-6361/201322312}, \href
  {https://ui.adsabs.harvard.edu/abs/2014A&A...564A..66B} {564, A66}

\bibitem[\protect\citeauthoryear{{Cora} et~al.,}{{Cora}
  et~al.}{2018}]{Cora:2018}
{Cora} S.~A.,  et~al., 2018, \mn@doi [\mnras] {10.1093/mnras/sty1131}, \href
  {http://adsabs.harvard.edu/abs/2018MNRAS.479....2C} {479, 2}

\bibitem[\protect\citeauthoryear{{Delfino}, {Sc{\'o}ccola}, {Cora},
  {Vega-Mart{\'\i}nez}  \& {Gargiulo}}{{Delfino} et~al.}{2022}]{Delfino2022}
{Delfino} F.~M.,  {Sc{\'o}ccola} C.~G.,  {Cora} S.~A.,  {Vega-Mart{\'\i}nez}
  C.~A.,   {Gargiulo} I.~D.,  2022, \mn@doi [\mnras] {10.1093/mnras/stab3494},
  \href {https://ui.adsabs.harvard.edu/abs/2022MNRAS.510.2900D} {510, 2900}

\bibitem[\protect\citeauthoryear{{Dressler}}{{Dressler}}{1980}]{Dressler:1980}
{Dressler} A.,  1980, \mn@doi [\apjs] {10.1086/190663}, \href
  {http://adsabs.harvard.edu/abs/1980ApJS...42..565D} {42, 565}

\bibitem[\protect\citeauthoryear{{Duplancic}, {D{\'a}vila-Kurb{\'a}n},
  {Coldwell}, {Alonso}  \& {Galdeano}}{{Duplancic}
  et~al.}{2020}]{Duplancic:2020}
{Duplancic} F.,  {D{\'a}vila-Kurb{\'a}n} F.,  {Coldwell} G.~V.,  {Alonso} S.,
  {Galdeano} D.,  2020, \mn@doi [\mnras] {10.1093/mnras/staa393}, \href
  {https://ui.adsabs.harvard.edu/abs/2020MNRAS.493.1818D} {493, 1818}

\bibitem[\protect\citeauthoryear{{Gruppioni} et~al.,}{{Gruppioni}
  et~al.}{2015}]{gruppioni15}
{Gruppioni} C.,  et~al., 2015, \mn@doi [\mnras] {10.1093/mnras/stv1204}, \href
  {http://adsabs.harvard.edu/abs/2015MNRAS.451.3419G} {451, 3419}

\bibitem[\protect\citeauthoryear{{Hahn}, {Porciani}, {Carollo}  \&
  {Dekel}}{{Hahn} et~al.}{2007a}]{Hahn:2007a}
{Hahn} O.,  {Porciani} C.,  {Carollo} C.~M.,   {Dekel} A.,  2007a, \mn@doi
  [\mnras] {10.1111/j.1365-2966.2006.11318.x}, \href
  {https://ui.adsabs.harvard.edu/abs/2007MNRAS.375..489H} {375, 489}

\bibitem[\protect\citeauthoryear{{Hahn}, {Carollo}, {Porciani}  \&
  {Dekel}}{{Hahn} et~al.}{2007b}]{Hahn:2007b}
{Hahn} O.,  {Carollo} C.~M.,  {Porciani} C.,   {Dekel} A.,  2007b, \mn@doi
  [\mnras] {10.1111/j.1365-2966.2007.12249.x}, \href
  {https://ui.adsabs.harvard.edu/abs/2007MNRAS.381...41H} {381, 41}

\bibitem[\protect\citeauthoryear{{Henriques}, {White}, {Thomas}, {Angulo},
  {Guo}, {Lemson}, {Springel}  \& {Overzier}}{{Henriques}
  et~al.}{2015}]{henriques_mcmc_2015}
{Henriques} B.~M.~B.,  {White} S.~D.~M.,  {Thomas} P.~A.,  {Angulo} R.,  {Guo}
  Q.,  {Lemson} G.,  {Springel} V.,   {Overzier} R.,  2015, \mn@doi [\mnras]
  {10.1093/mnras/stv705}, \href
  {http://adsabs.harvard.edu/abs/2015MNRAS.451.2663H} {451, 2663}

\bibitem[\protect\citeauthoryear{{Hoffman}, {Metuki}, {Yepes}, {Gottl{\"o}ber},
  {Forero-Romero}, {Libeskind}  \& {Knebe}}{{Hoffman}
  et~al.}{2012}]{Hoffman2012}
{Hoffman} Y.,  {Metuki} O.,  {Yepes} G.,  {Gottl{\"o}ber} S.,  {Forero-Romero}
  J.~E.,  {Libeskind} N.~I.,   {Knebe} A.,  2012, \mn@doi [\mnras]
  {10.1111/j.1365-2966.2012.21553.x}, \href
  {https://ui.adsabs.harvard.edu/abs/2012MNRAS.425.2049H} {425, 2049}

\bibitem[\protect\citeauthoryear{{Huchra} \& {Geller}}{{Huchra} \&
  {Geller}}{1982}]{H&G:1982}
{Huchra} J.~P.,  {Geller} M.~J.,  1982, \mn@doi [\apj] {10.1086/160000}, \href
  {http://adsabs.harvard.edu/abs/1982ApJ...257..423H} {257, 423}

\bibitem[\protect\citeauthoryear{{Ivezi{\'c}} et~al.,}{{Ivezi{\'c}}
  et~al.}{2019}]{LSST:2019}
{Ivezi{\'c}} {\v{Z}}.,  et~al., 2019, \mn@doi [\apj]
  {10.3847/1538-4357/ab042c}, \href
  {https://ui.adsabs.harvard.edu/abs/2019ApJ...873..111I} {873, 111}

\bibitem[\protect\citeauthoryear{{Kauffmann}, {White}, {Heckman}, {M{\'e}nard},
  {Brinchmann}, {Charlot}, {Tremonti}  \& {Brinkmann}}{{Kauffmann}
  et~al.}{2004}]{Kauffmann:2004}
{Kauffmann} G.,  {White} S.~D.~M.,  {Heckman} T.~M.,  {M{\'e}nard} B.,
  {Brinchmann} J.,  {Charlot} S.,  {Tremonti} C.,   {Brinkmann} J.,  2004,
  \mn@doi [\mnras] {10.1111/j.1365-2966.2004.08117.x}, \href
  {http://adsabs.harvard.edu/abs/2004MNRAS.353..713K} {353, 713}

\bibitem[\protect\citeauthoryear{{Klypin}, {Yepes}, {Gottl{\"o}ber}, {Prada}
  \& {He{\ss}}}{{Klypin} et~al.}{2016}]{Klypin:2016}
{Klypin} A.,  {Yepes} G.,  {Gottl{\"o}ber} S.,  {Prada} F.,   {He{\ss}} S.,
  2016, \mn@doi [\mnras] {10.1093/mnras/stw248}, \href
  {http://adsabs.harvard.edu/abs/2016MNRAS.457.4340K} {457, 4340}

\bibitem[\protect\citeauthoryear{{Knebe} et~al.,}{{Knebe}
  et~al.}{2018}]{Knebe:2018}
{Knebe} A.,  et~al., 2018, \mn@doi [\mnras] {10.1093/mnras/stx3274}, \href
  {https://ui.adsabs.harvard.edu/abs/2018MNRAS.475.2936K} {475, 2936}

\bibitem[\protect\citeauthoryear{{Kormendy} \& {Ho}}{{Kormendy} \&
  {Ho}}{2013}]{kormendy_bhb_2013}
{Kormendy} J.,  {Ho} L.~C.,  2013, \mn@doi [\araa]
  {10.1146/annurev-astro-082708-101811}, \href
  {https://ui.adsabs.harvard.edu/abs/2013ARA&A..51..511K} {51, 511}

\bibitem[\protect\citeauthoryear{{Libeskind}, {Hoffman}, {Forero-Romero},
  {Gottl{\"o}ber}, {Knebe}, {Steinmetz}  \& {Klypin}}{{Libeskind}
  et~al.}{2013}]{Libeskind2013}
{Libeskind} N.~I.,  {Hoffman} Y.,  {Forero-Romero} J.,  {Gottl{\"o}ber} S.,
  {Knebe} A.,  {Steinmetz} M.,   {Klypin} A.,  2013, \mn@doi [\mnras]
  {10.1093/mnras/sts216}, \href
  {https://ui.adsabs.harvard.edu/abs/2013MNRAS.428.2489L} {428, 2489}

\bibitem[\protect\citeauthoryear{{Libeskind} et~al.,}{{Libeskind}
  et~al.}{2018}]{Libeskind2018}
{Libeskind} N.~I.,  et~al., 2018, \mn@doi [\mnras] {10.1093/mnras/stx1976},
  \href {https://ui.adsabs.harvard.edu/abs/2018MNRAS.473.1195L} {473, 1195}

\bibitem[\protect\citeauthoryear{{McConnell} \& {Ma}}{{McConnell} \&
  {Ma}}{2013}]{mcconnell_bhb_2013}
{McConnell} N.~J.,  {Ma} C.-P.,  2013, \mn@doi [\apj]
  {10.1088/0004-637X/764/2/184}, \href
  {https://ui.adsabs.harvard.edu/abs/2013ApJ...764..184M} {764, 184}

\bibitem[\protect\citeauthoryear{{Muratov}, {Kere{\v{s}}},
  {Faucher-Gigu{\`e}re}, {Hopkins}, {Quataert}  \& {Murray}}{{Muratov}
  et~al.}{2015}]{Muratov:2015}
{Muratov} A.~L.,  {Kere{\v{s}}} D.,  {Faucher-Gigu{\`e}re} C.-A.,  {Hopkins}
  P.~F.,  {Quataert} E.,   {Murray} N.,  2015, \mn@doi [\mnras]
  {10.1093/mnras/stv2126}, \href
  {https://ui.adsabs.harvard.edu/abs/2015MNRAS.454.2691M} {454, 2691}

\bibitem[\protect\citeauthoryear{{O'Mill}, {Padilla}  \& {Garc{\'\i}a
  Lambas}}{{O'Mill} et~al.}{2008}]{OMill:2008}
{O'Mill} A.~L.,  {Padilla} N.,   {Garc{\'\i}a Lambas} D.,  2008, \mn@doi
  [\mnras] {10.1111/j.1365-2966.2008.13650.x}, \href
  {https://ui.adsabs.harvard.edu/abs/2008MNRAS.389.1763O} {389, 1763}

\bibitem[\protect\citeauthoryear{{Peng} et~al.,}{{Peng}
  et~al.}{2010}]{Peng:2010}
{Peng} Y.-j.,  et~al., 2010, \mn@doi [\apj] {10.1088/0004-637X/721/1/193},
  \href {http://adsabs.harvard.edu/abs/2010ApJ...721..193P} {721, 193}

\bibitem[\protect\citeauthoryear{{Planck Collaboration} et~al.,}{{Planck
  Collaboration} et~al.}{2014}]{Planck:2014}
{Planck Collaboration} et~al., 2014, \mn@doi [\aap]
  {10.1051/0004-6361/201321591}, \href
  {http://adsabs.harvard.edu/abs/2014A%26A...571A..16P} {571, A16}

\bibitem[\protect\citeauthoryear{{Porciani}, {Dekel}  \& {Hoffman}}{{Porciani}
  et~al.}{2002}]{Porciani2002}
{Porciani} C.,  {Dekel} A.,   {Hoffman} Y.,  2002, \mn@doi [\mnras]
  {10.1046/j.1365-8711.2002.05306.x}, \href
  {https://ui.adsabs.harvard.edu/abs/2002MNRAS.332..339P} {332, 339}

\bibitem[\protect\citeauthoryear{{Ruiz} et~al.,}{{Ruiz}
  et~al.}{2015}]{Ruiz:2015}
{Ruiz} A.~N.,  et~al., 2015, \mn@doi [\apj] {10.1088/0004-637X/801/2/139},
  \href {http://adsabs.harvard.edu/abs/2015ApJ...801..139R} {801, 139}

\bibitem[\protect\citeauthoryear{{Ruiz}, {Alfaro}  \& {Garcia Lambas}}{{Ruiz}
  et~al.}{2019}]{Ruiz:2019}
{Ruiz} A.~N.,  {Alfaro} I.~G.,   {Garcia Lambas} D.,  2019, \mn@doi [\mnras]
  {10.1093/mnras/sty3443}, \href
  {https://ui.adsabs.harvard.edu/abs/2019MNRAS.483.4070R} {483, 4070}

\bibitem[\protect\citeauthoryear{{Springel}, {White}, {Tormen}  \&
  {Kauffmann}}{{Springel} et~al.}{2001}]{Springel:2001}
{Springel} V.,  {White} S.~D.~M.,  {Tormen} G.,   {Kauffmann} G.,  2001,
  \mn@doi [\mnras] {10.1046/j.1365-8711.2001.04912.x}, \href
  {http://adsabs.harvard.edu/abs/2001MNRAS.328..726S} {328, 726}

\bibitem[\protect\citeauthoryear{{Taverna}, {Salerno}, {Daza-Perilla},
  {D{\'\i}az-Gim{\'e}nez}, {Zandivarez}, {Mart{\'\i}nez}  \& {Ruiz}}{{Taverna}
  et~al.}{2023}]{Taverna:2023}
{Taverna} A.,  {Salerno} J.~M.,  {Daza-Perilla} I.~V.,  {D{\'\i}az-Gim{\'e}nez}
  E.,  {Zandivarez} A.,  {Mart{\'\i}nez} H.~J.,   {Ruiz} A.~N.,  2023, \mn@doi
  [\mnras] {10.1093/mnras/stad416}, \href
  {https://ui.adsabs.harvard.edu/abs/2023MNRAS.520.6367T} {520, 6367}

\bibitem[\protect\citeauthoryear{{Tempel}, {Tago}  \& {Liivam{\"a}gi}}{{Tempel}
  et~al.}{2012}]{Tempel:2012}
{Tempel} E.,  {Tago} E.,   {Liivam{\"a}gi} L.~J.,  2012, \mn@doi [\aap]
  {10.1051/0004-6361/201118687}, \href
  {https://ui.adsabs.harvard.edu/abs/2012A&A...540A.106T} {540, A106}

\bibitem[\protect\citeauthoryear{{Tully}}{{Tully}}{1987}]{Tully:1987}
{Tully} R.~B.,  1987, \mn@doi [\apj] {10.1086/165629}, \href
  {http://adsabs.harvard.edu/abs/1987ApJ...321..280T} {321, 280}

\bibitem[\protect\citeauthoryear{{Tully}}{{Tully}}{1988}]{Tully:1988}
{Tully} R.~B.,  1988, Science, \href
  {http://adsabs.harvard.edu/abs/1988Sci...242..310T} {242, 310}

\bibitem[\protect\citeauthoryear{{Tully} et~al.,}{{Tully}
  et~al.}{2006}]{Tully:2006}
{Tully} R.~B.,  et~al., 2006, \mn@doi [\aj] {10.1086/505466}, \href
  {http://adsabs.harvard.edu/abs/2006AJ....132..729T} {132, 729}

\bibitem[\protect\citeauthoryear{{Wang} et~al.,}{{Wang}
  et~al.}{2018}]{Wang2018}
{Wang} H.,  et~al., 2018, \mn@doi [\apj] {10.3847/1538-4357/aa9e01}, \href
  {https://ui.adsabs.harvard.edu/abs/2018ApJ...852...31W} {852, 31}

\bibitem[\protect\citeauthoryear{{Wang}, {Kang}, {Libeskind}, {Guo},
  {Gottl{\"o}ber}  \& {Wang}}{{Wang} et~al.}{2020}]{Wang:2020}
{Wang} P.,  {Kang} X.,  {Libeskind} N.~I.,  {Guo} Q.,  {Gottl{\"o}ber} S.,
  {Wang} W.,  2020, \mn@doi [\na] {10.1016/j.newast.2020.101405}, \href
  {https://ui.adsabs.harvard.edu/abs/2020NewA...8001405W} {80, 101405}

\bibitem[\protect\citeauthoryear{{Wetzel}, {Tinker}  \& {Conroy}}{{Wetzel}
  et~al.}{2012}]{Wetzel2012}
{Wetzel} A.~R.,  {Tinker} J.~L.,   {Conroy} C.,  2012, \mn@doi [\mnras]
  {10.1111/j.1365-2966.2012.21188.x}, \href
  {https://ui.adsabs.harvard.edu/abs/2012MNRAS.424..232W} {424, 232}

\bibitem[\protect\citeauthoryear{{Yang}, {Mo}, {van den Bosch}, {Pasquali},
  {Li}  \& {Barden}}{{Yang} et~al.}{2007}]{Yang:2007}
{Yang} X.,  {Mo} H.~J.,  {van den Bosch} F.~C.,  {Pasquali} A.,  {Li} C.,
  {Barden} M.,  2007, \mn@doi [\apj] {10.1086/522027}, \href
  {https://ui.adsabs.harvard.edu/abs/2007ApJ...671..153Y} {671, 153}

\bibitem[\protect\citeauthoryear{{Yaryura} et~al.,}{{Yaryura}
  et~al.}{2020}]{Yaryura:2020}
{Yaryura} C.~Y.,  et~al., 2020, \mn@doi [\mnras] {10.1093/mnras/staa3197},
  \href {https://ui.adsabs.harvard.edu/abs/2020MNRAS.499.5932Y} {499, 5932}

\bibitem[\protect\citeauthoryear{{Zhang}, {Yang}, {Faltenbacher}, {Springel},
  {Lin}  \& {Wang}}{{Zhang} et~al.}{2009}]{Zhang:2009}
{Zhang} Y.,  {Yang} X.,  {Faltenbacher} A.,  {Springel} V.,  {Lin} W.,   {Wang}
  H.,  2009, \mn@doi [\apj] {10.1088/0004-637X/706/1/747}, \href
  {https://ui.adsabs.harvard.edu/abs/2009ApJ...706..747Z} {706, 747}

\bibitem[\protect\citeauthoryear{{Zheng} et~al.,}{{Zheng}
  et~al.}{2017}]{Zheng:2017}
{Zheng} Z.,  et~al., 2017, \mn@doi [\mnras] {10.1093/mnras/stw3030}, \href
  {https://ui.adsabs.harvard.edu/abs/2017MNRAS.465.4572Z} {465, 4572}

\bibitem[\protect\citeauthoryear{{de Jong} et~al.,}{{de Jong}
  et~al.}{2019}]{deJong:2019}
{de Jong} R.~S.,  et~al., 2019, \mn@doi [The Messenger]
  {10.18727/0722-6691/5117}, \href
  {https://ui.adsabs.harvard.edu/abs/2019Msngr.175....3D} {175, 3}

\makeatother
\end{thebibliography}

\appendix
\section{Stability of classification of environment}
\label{S_appendix}

\begin{table}
\centering
   \begin{tabular}{  c  c  c  c  }
    \hline
    \hline
    \\
       &\multicolumn{3}{c}{TIDAL}  
    \\ 
    \hline
     & $l=1$ & $l=2$ & $l=4$ 
    \\
    \hline
    KNOTS & 17671 & 28087 & 34667   
    \\
    FILAMENTS & 206839 & 182974 & 165609   
    \\
    WALLS & 81407 & 91564 & 98662   
     \\
    VOIDS & 2333 & 5625 & 9312   
     \\
    \hline
    \hline
  \end{tabular}
  \caption{Number of dwarf galaxy associations in different environments 
  (knots, filaments, walls, voids) estimated from the tidal tensor 
  for three different values of smoothing length, $l = 1, 2$ and 4 
  $\,{\rm Mpc}\,h^{-1}$.}
 \label{table:table1}
\end{table}

\begin{figure}
    \begin{subfigure}{
        \includegraphics[width=0.24\textwidth]{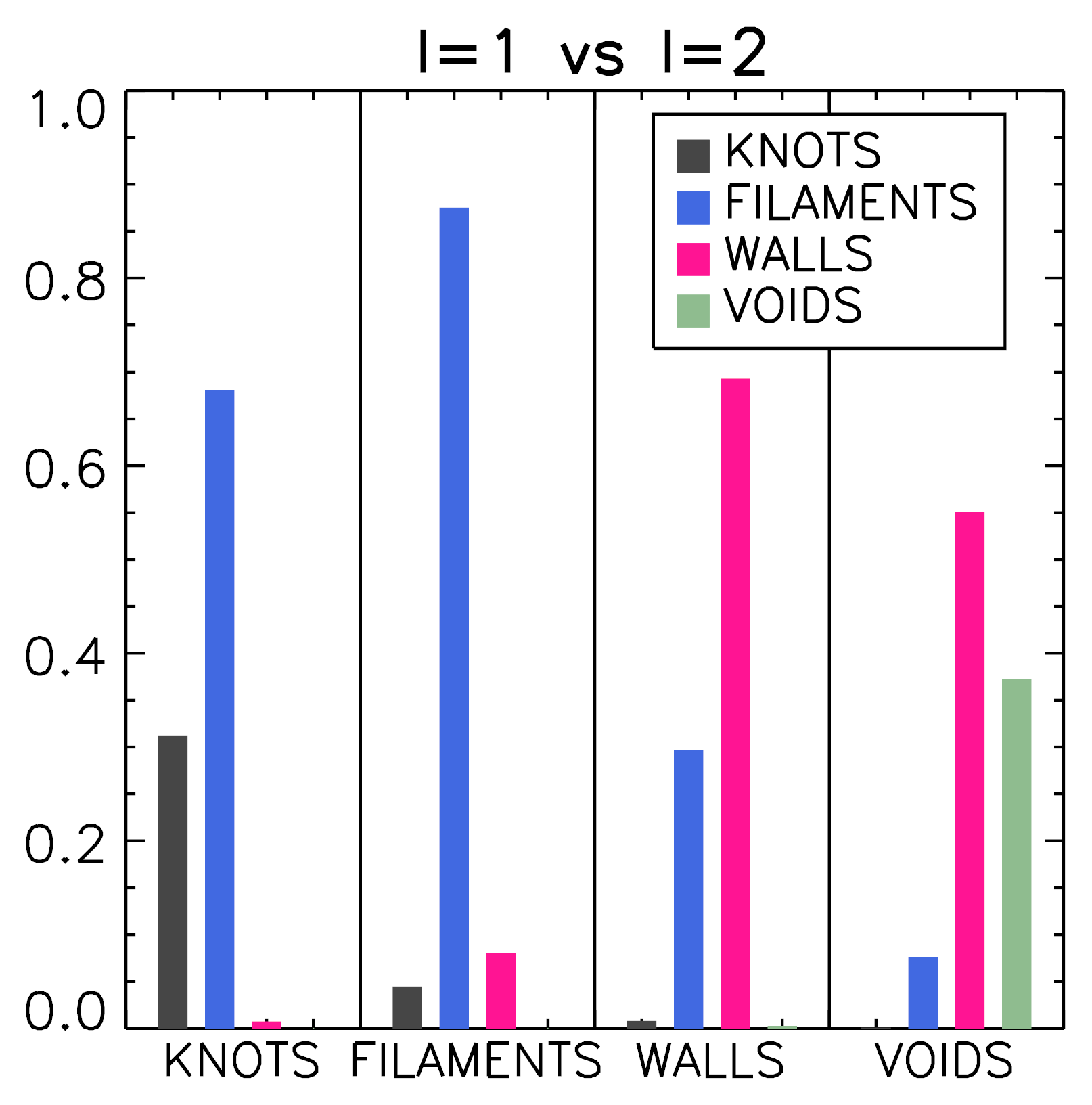}
        \includegraphics[width=0.24\textwidth]{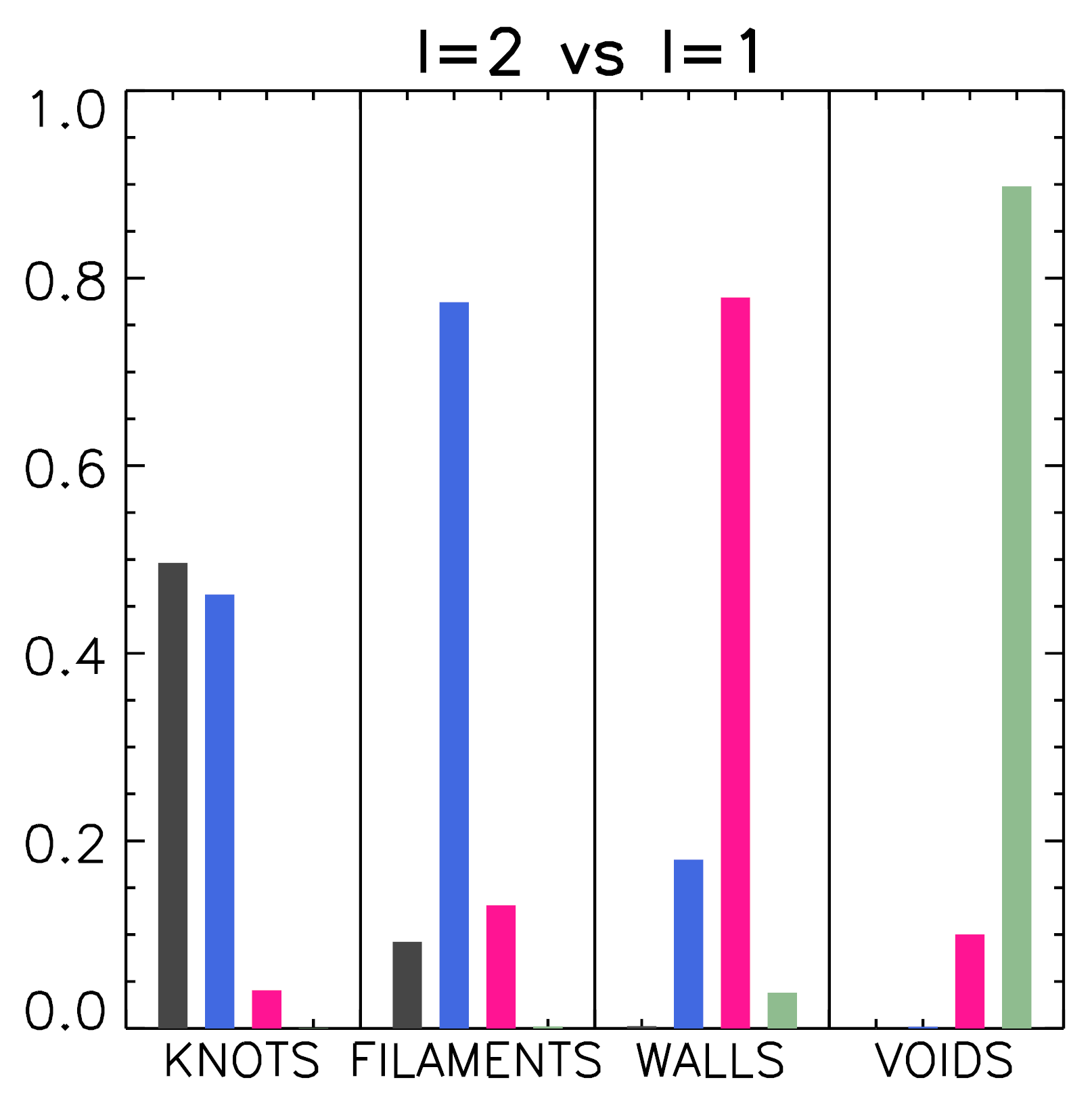}}
    \end{subfigure}
    \begin{subfigure}{
        \includegraphics[width=0.24\textwidth]{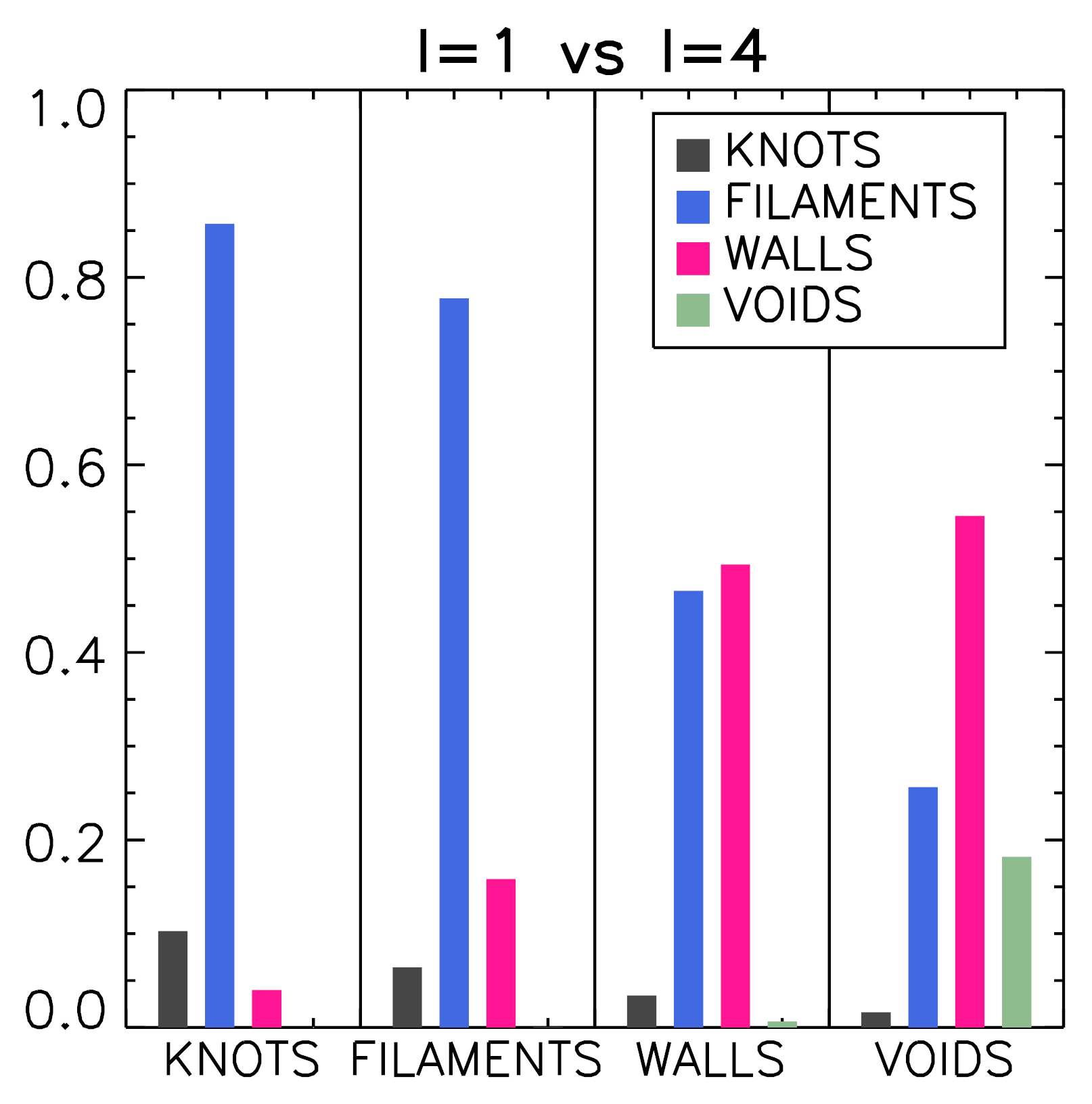}
        \includegraphics[width=0.24\textwidth]{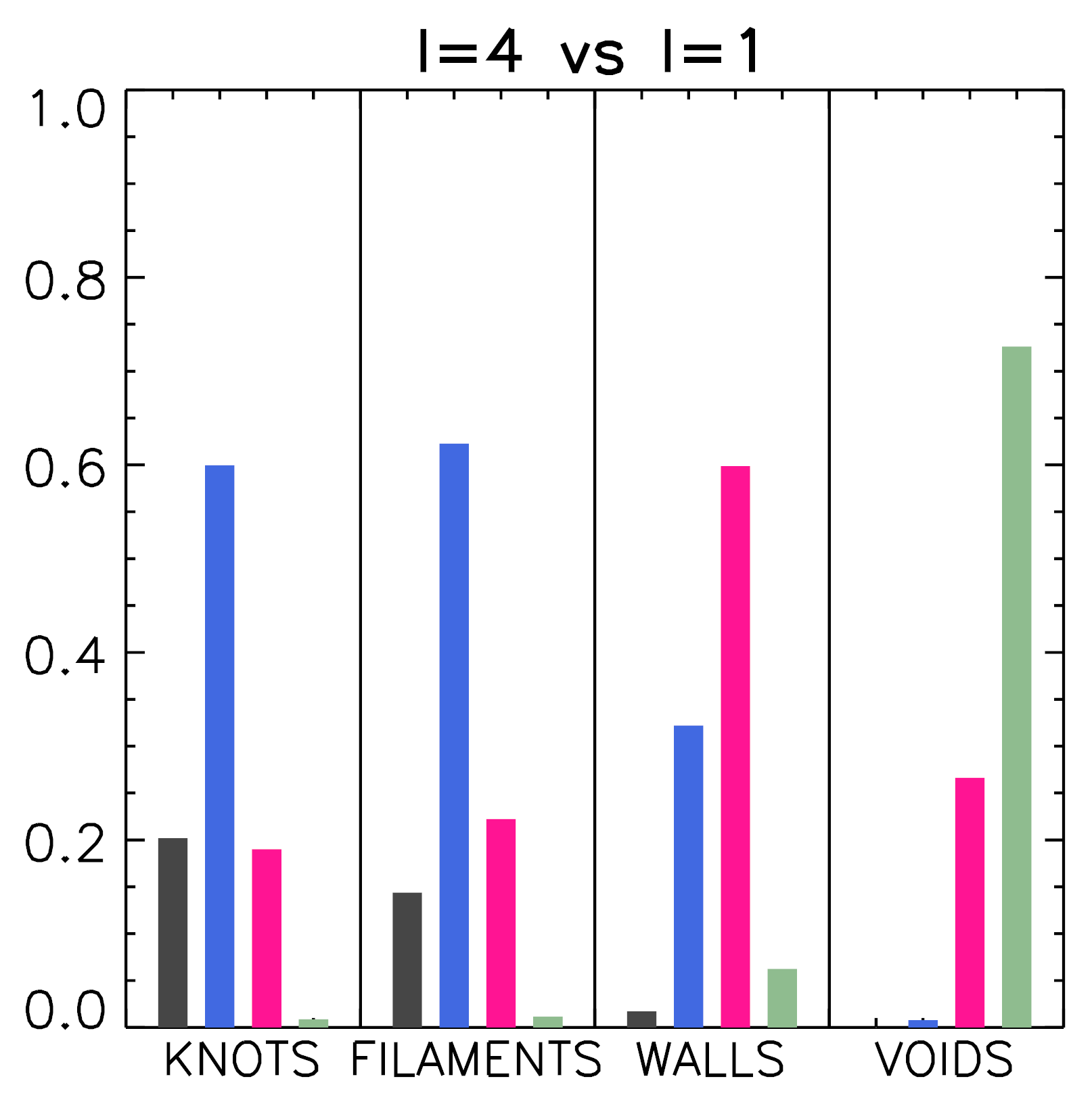}}
    \end{subfigure}
    \begin{subfigure}{
        \includegraphics[width=0.24\textwidth]{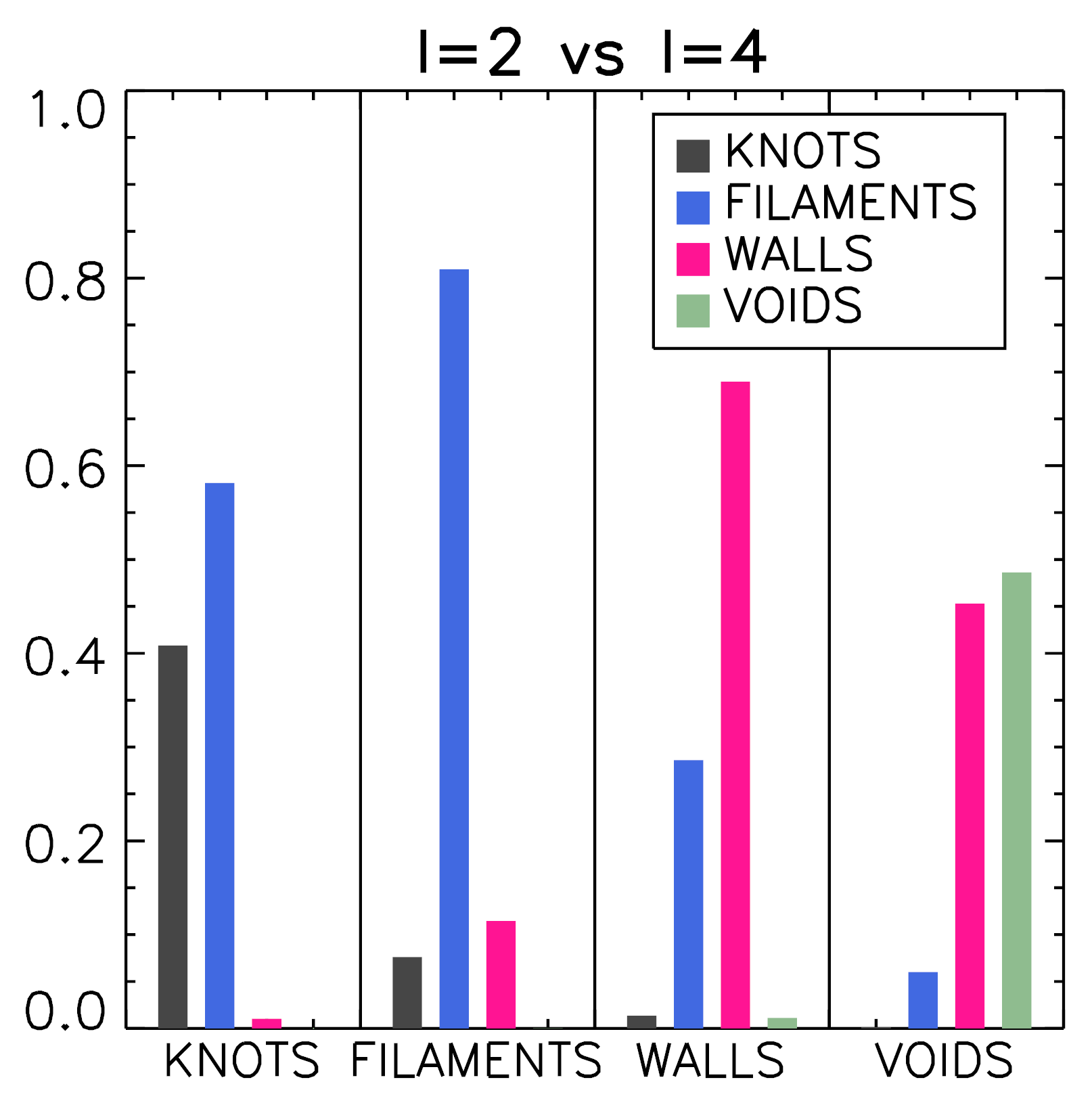}
        \includegraphics[width=0.24\textwidth]{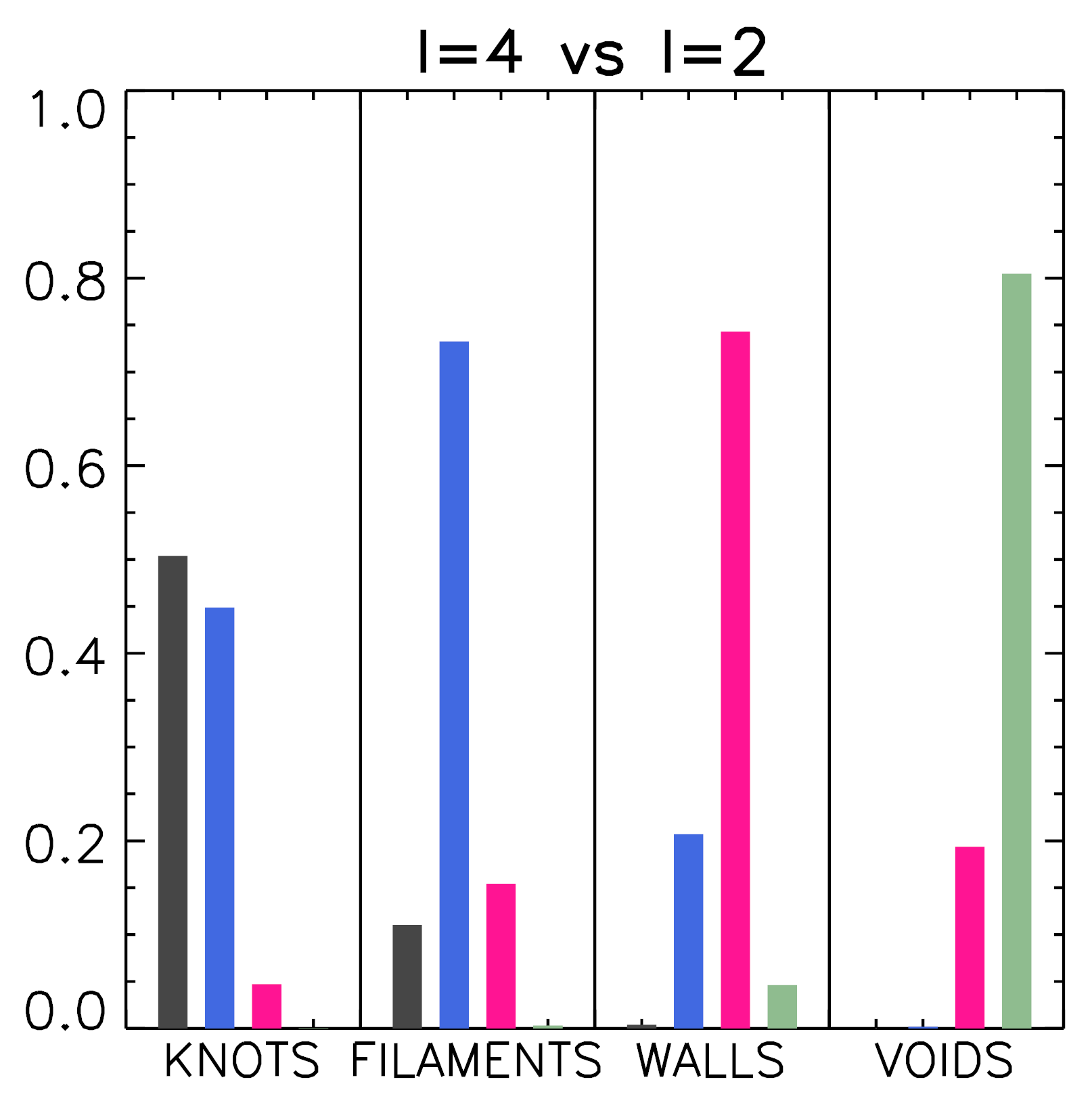}}
    \end{subfigure}
    \caption{Change of associations from one category (knot, filament, wall or void) 
    to another according to the smoothing length used to estimate the eigenvalues, 
    comparing 2 samples at a time. 
    Upper, middle and lower panels show the comparison 
    between $l=1$ vs. $l=2$, $l=1$ vs. $l=4$ and  $l=2$ vs. $l=4$ 
    and their reciprocals, respectively, as indicated by legends. }
    \label{fig:appe}
\end{figure}

In this appendix, we analyse the robustness of our 
classification scheme
of the large–scale matter distribution of the cosmic web
according to the variation of the 
characteristic scale adopted to estimate the density 
of the environment.
In particular, we evaluate how the number of associations 
categorised in a given environment (knot, filament, wall or void) 
varies when changing the smoothing length 
($l = 1, 2$ or 4 $\,{\rm Mpc}\,h^{-1}$). 
These four categories are assigned according to the sign of the 
eigenvalues of the Hessian matrix estimated from the 
gravitational potential (see Section \ref{S_kfwv}).
To compute these eigenvalues, first, the full simulated box of 
the \smdpl~simulation (described in Section \ref{sec:simSMDPL}) 
is covered with a fixed grid of $400^3$ cells.
Then, the DM particle overdensity is smoothed using a spherically 
symmetric Gaussian window function with a given $l$, and 
the gravitational potential is estimated. 
The gravitational potential ($\phi$) is normalised by 
$4 \pi G \overline{\rho}$ whereby it satisfies the Poisson 
equation ($\bigtriangledown ^{2} \phi = \delta$), where $\delta$ is 
the dimensionless matter overdensity, $G$ the gravitational 
constant and $\overline{\rho}$ is the average density of 
the Universe.
The Hessian matrix is constructed from this peculiar 
gravitational potential (tidal tensor) and we assign its three 
eigenvalues to each cell. 
Finally, we assign to each association the same eigenvalues 
as those of the cell in which the association centre is located.
As expected, these eigenvalues depend on the $l$ of the Gaussian 
filter. 
For a given $l$, the number of eigenvalues greater than a given 
threshold ($\lambda_{\rm th} = 0$) is used to 
classify the type of environment where each association resides: 
knot, filament, wall or void. 
Table \ref{table:table1} presents the number of dwarf galaxy 
associations in each environment for the three different 
smoothing lengths.
\\

Fig.~\ref{fig:appe} shows how systems are identified as part of
one environment (knot, filament, wall or void) or
another according to the $l$ used.
The upper, middle and lower panels show the comparison between 
$l=1$ vs. $l=2$, $l=1$ vs. $l=4$ and $l=2$ vs. $l=4$ 
and their reciprocals, 
respectively, as indicated by the legends. 
On the x-axis of each panel, there are four boxes
corresponding to knots, filaments, walls and voids.
In each of the boxes, there are four bars with different colours 
identifying knots (black), filaments (blue), 
walls (magenta) and voids (green). 
The length of each bar corresponds to the ratio 
$N_{XY}/N_{Y}$, where $N_{XY}$ is the number of objects 
belonging to category $X$ in, for example, $l=1$ but categorised 
as $Y$ in, for example, $l=2$,
while $N_Y$ is the total number of associations categorised as $Y$ 
in $l=2$ following the same example as above, 
where $X$ and $Y$ can take the values K, F, W or V,
for knot, filament, wall and void, respectively.
The mentioned example corresponds to $l=1$ vs. $l=2$, 
shown in the upper left panel of Fig.~\ref{fig:appe}. 
Notice that with this definition it is true that

\begin{equation}
  N_{Y} = N_{{\rm K}Y} + N_{{\rm F}Y} + N_{{\rm W}Y} + N_{{\rm V}Y},
\end{equation}

By definition the sum of the length of the bars in each box 
is equal to 1, so the different lengths give an idea about the `wandering' 
of objects from one environment to another of the cosmic web.
If the bar length equals 1, all objects stay in the same environment.
So, if the categorisation does not change from one smoothing length to another, 
the first box of each panel ($Y={\rm K}$) should show only a black bar, 
the second box ($Y={\rm F}$) should show only a blue bar,
the third box ($Y={\rm W}$) should show only a magenta bar and 
the fourth box ($Y={\rm V}$) should show only a green bar, 
all of which should have the same length (equal to 1).
Evidently, this is not what we see in Fig.~\ref{fig:appe}.
Filaments and walls seem to be the best defined environments.
The classification of walls is a bit less stable, in particular, 
considering the large step from 1 to 4 $\,{\rm Mpc}\,h^{-1}$.
The right panels show that decreasing the smoothing length 
keeps the objects in the same environment. 
The classification of filaments seems to be the most stable,
while the classification of knots is the most unstable for all cases.
\\

Despite the variations described in the different categories when we classify the 
environment considering different smoothing length values, the results presented in 
this work are in full agreement for the three chosen smoothing lengths. 
Therefore, we decided to present only results based on the smoothing length 1 Mpc/h. 
We have also tested the stability of the environment definition based on tidal tensor 
versus shear velocity. 
We find an eighty to ninety percent agreement for filaments, walls and voids. 
Again, the main results presented in this project are in full agreement if we use 
the shear velocity field instead of the tidal tensor.


\bsp	
\label{lastpage}

\end{document}